\documentstyle[aps,twocolumn,epsf,twocolumn,eqsecnum,floats]{revtex}
\begin{document}
\draft
\wideabs{
\title{
Seismic gravity-gradient noise in interferometric gravitational-wave
detectors
}
\author{Scott A.\ Hughes$^{(1)}$ and Kip S.\ Thorne$^{(1,2)}$} 
\address{
$^{(1)}$Theoretical Astrophysics, California Institute of Technology,
Pasadena, CA 91125}
\address{
$^{(2)}$Max-Planck-Institut f\"ur Gravitationsphysik, Schlatzweg 1,
14473 Potsdam, Germany}
\maketitle
\begin{abstract}
When ambient seismic waves pass near and under an interferometric
gravitational-wave detector, they induce density perturbations in the
earth, which in turn produce fluctuating gravitational forces on the
interferometer's test masses.  These forces mimic a stochastic
background of gravitational waves and thus constitute a noise source.
This {{\em seismic gravity-gradient noise}} has been estimated and
discussed previously by Saulson using a simple model of the earth's
ambient seismic motions.  In this paper, we develop a more
sophisticated model of these motions, based on the theory of multimode
Rayleigh and Love waves propagating in a multilayer medium that
approximates the geological strata at the LIGO sites (Tables
{\ref{table:skagit}}--{\ref{table:4LayerLivingston}}), and we use this
model to revisit seismic gravity gradients.  We characterize the
seismic gravity-gradient noise by a transfer function,
$T(f)\equiv{\tilde x}(f)/{\tilde W}(f)$, from the spectrum of rms
seismic displacements averaged over vertical and horizontal
directions, ${\tilde W}(f)$, to the spectrum of interferometric
test-mass motions, ${\tilde x}(f) \equiv L \tilde h (f)$; here $L$ is
the interferometer arm length, $\tilde h(f)$ is the gravitational-wave
noise spectrum, and $f$ is frequency.  Our model predicts a transfer
function with essentially the same functional form as that derived by
Saulson, $T \simeq 4\pi G\rho(2\pi f)^{-2}\beta(f)$, where $\rho$ is
the density of the earth near the test masses, $G$ is Newton's
constant, and $\beta(f) \equiv \gamma(f) \Gamma(f)\beta'(f)$ is a
dimensionless {\it reduced transfer function} whose components $\gamma
\simeq 1$ and $\Gamma \simeq 1$ account for a weak correlation between
the interferometer's two test masses (Fig.\ \ref{fig:gamma}) and a
slight reduction of the noise due to the height of the test masses
above the earth's surface.  This paper's primary foci are (i) a study
of how $\beta'(f) \simeq \beta(f)$ depends on the various Rayleigh and
Love modes that are present in the seismic spectrum (Figs.\
{\ref{fig:betapUp}}--{\ref{fig:DRLivingstonP}} and Table
{\ref{table:betap}}), (ii) an attempt to estimate which modes are
actually present at the two LIGO sites at quiet times and at noisy
times, and (iii) a corresponding estimate of the magnitude of
$\beta'(f)$ at quiet and noisy times.  We conclude that {\it at quiet
times} $\beta' \simeq 0.35$--0.6 at the LIGO sites, and {\it at noisy
times} $\beta' \simeq 0.15$--1.4. (For comparison, Saulson's simple
model gave $\beta = \beta' = 1/\sqrt3 = 0.58$.)  By folding our
resulting transfer function into the ``standard LIGO seismic
spectrum'' {[Eq.\ (\ref{eq:hSGG})]}, which approximates $\tilde W(f)$
at typical times, we obtain the gravity-gradient noise spectra shown
in Fig.\ \ref{fig:hSGG}.  At quiet times this noise is below the
benchmark noise level of ``advanced LIGO interferometers'' at all
frequencies (though not by much at $\sim 10$ Hz); at noisy times it
may significantly exceed the advanced noise level near 10 Hz.  The
lower edge of our quiet-time noise constitutes a limit, beyond which
there would be little gain from further improvements in vibration
isolation and thermal noise, unless one can also reduce the seismic
gravity gradient noise.  Two methods of such reduction are briefly
discussed: monitoring the earth's density perturbations near each test
mass, computing the gravitational forces they produce, and correcting
the data for those forces; and constructing narrow moats around the
interferometers' corner and end stations to shield out the
fundamental-mode Rayleigh waves, which we suspect dominate at quiet
times.
\end{abstract}
\pacs{PACS numbers: 04.80.Nn}
}

\narrowtext

\section{Introduction and Summary}

Now that the LIGO/VIRGO international network of gravitational-wave
detectors {\cite{abramovici,virgo,snowmass_summary,snowmass_kip}} is
under construction, it is important to revisit the various noise
sources that will constrain the network's ultimate performance.
Improved estimates of the ultimate noise spectra are a foundation for
long-term planning on a number of aspects of gravitational-wave
research, including facilities design, interferometer R\&D, data
analysis algorithm development, and astrophysical source studies.

In this paper and a subsequent one {\cite{winstein_thorne}} we revisit
{\it gravity-gradient noise} --- noise due to fluctuating Newtonian
gravitational forces that induce motions in the test masses of an
interferometric gravitational-wave detector.  Gravity gradients are
potentially important at the low end of the interferometers' frequency
range, $f\alt 20$ Hz.  Another noise source that is important at these
frequencies is {\it vibrational seismic noise}, in which the ground's
ambient motions, filtered through the detector's vibration isolation
system, produce motions of the test masses.  It should be possible and
practical to isolate the test masses from these seismic vibrations
down to frequencies as low as $f\sim 3$ Hz {\cite{seismic_isolation}},
but it does not look practical to achieve large amounts of isolation
from the fluctuating gravity gradients.  Thus, gravity gradients
constitute an ultimate low-frequency noise source; seismic vibrations
do not.

Gravity gradients were first identified as a potential noise source in
interferometric gravitational-wave detectors by Rai Weiss in 1972
{\cite{weiss72}}.  The first quantitative analyses of such
gravity-gradient noise were performed by Peter Saulson
{\cite{saulson83}} and Robert Spero {\cite{spero82}} in the early
1980s.  There has been little further study of gravity-gradient noise
since then, except for some updating in Saulson's recent monograph
{\cite{saulson_monograph}}.

In his updating, Saulson concluded that the most serious source of
gravity-gradient noise will be the fluctuating density of the earth
beneath and near each of the interferometer's test masses.  These
density fluctuations are induced by ambient seismic waves that are
always present; their resulting gravitational forces are called {\it
seismic gravity-gradient noise}.  Saulson
{\cite{saulson83,saulson_monograph}} also estimated the gravity
gradient noise from atmospheric fluctuations, concluding that it is
probably weaker than that from earth motions.  Spero {\cite{spero82}}
showed that gravity-gradient noise due to jerky human activity (and
that of dogs, cattle, and other moving bodies) can be more serious
than seismic gravity-gradient noise if such bodies are not kept at an
adequate distance from the test masses.  We shall revisit seismic
gravity-gradient noise in this paper, and gravity gradients due to
human activity in a subsequent one {\cite{winstein_thorne}}; Teviet
Creighton at Caltech has recently initiated a careful revisit of
gravity gradient noise due to atmospheric fluctuations.

Our detailed analysis in this paper reveals a level of seismic
gravity-gradient noise that agrees remarkably well with Saulson's much
cruder estimates.  Our analysis reveals the uncertainties in the
gravity gradient noise, the range in which the noise may vary from
seismically quiet times to noisy times, the dependence of the noise on
the various seismic modes that are excited, and the characteristics of
the modes that the geological strata at Hanford and Livingston are
likely to support.  The dependence of the noise on the modes and the
characteristics of the expected modes are potential foundations for
methods of mitigating the seismic gravity gradient noise, discussed in
our concluding section.

A preliminary version of this paper {\cite{hughes_thorne}} was
circulated to the gravitational-wave-detection community in 1996.
That version considered only fundamental-mode Rayleigh waves (which we
suspect are responsible for the dominant seismic gravity-gradient
noise at quiet times), and (as Ken Libbrecht pointed out to us) it
contained a serious error: the omission of the ``surface-source'' term
[denoted $\xi_V$ in Eq.\ (\ref{eq:betapJ}) below] for the
gravity-gradient force. It also contained errors in its
two-geological-layer analysis for the LIGO Hanford site.  These errors
have been corrected in this final version of the manuscript, and the
analysis has been extended to include more realistic models of the
geological strata at the two LIGO sites and to include higher-order
seismic modes.

As we were completing this manuscript, we learned of a paper in press
{\cite{cella_cuoco}} by Giancarlo Cella, Elena Cuoco and their
VIRGO-Project collaborators, which also analyzes seismic
gravity-gradient noise in interferometric gravitational wave
detectors.  That paper is complementary to ours. Both papers analyze
the RF mode (which we suspect is the dominant contributor to the
seismic gravity-gradient noise at quiet times), obtaining the same
results in the 3--30 Hz band when the effects of geological
stratification are neglected.  But, whereas our paper carries out an
extensive study of the effects of stratification and other modes, the
Cella-Cuoco paper extends the unstratified RF-mode analysis to
frequencies below 3 Hz and above 30 Hz, and computes (and finds to be
small) the gravity gradient noise caused by seismically-induced
motions of the experimental apparatus and its massive physical
infrastructure in the vicinity of the VIRGO test masses.

This paper is organized as follows: In Sec.\
{\ref{sec:SitePhenomenology}}, we describe the phenomenology of the
seismic-wave modes that can contribute to ambient earth motions at
horizontally stratified sites like LIGO-Hanford and LIGO-Livingston.
In Sec.\ {\ref{sec:Tfunction}}, we introduce the transfer function
$T(f)$ used to characterize seismic gravity-gradient noise, we break
it down into its components [most especially the reduced transfer
function $\beta'(f)$], and we express it as an incoherent sum over
contributions from the various seismic modes.  In Sec.\
{\ref{sec:Tsaulson}}, we briefly describe Saulson's computation of the
reduced transfer function, and then in Sec.\ \ref{sec:Tus} we describe
our own computation and results.  More specifically, in \ref{sec:Tus}
we gather together and summarize from the body of the paper our
principal conclusions about $\beta'$ for the various modes at the two
LIGO sites, we discuss the evidence as to which modes actually
contribute to the noise at quiet times and at noisy times, and we
therefrom estimate the net values of $\beta'$ at quiet and noisy
times.  We then fold those estimates into the standard LIGO seismic
spectrum to get spectral estimates of the seismic gravity-gradient
noise (Fig.\
\ref{fig:hSGG}).

The remainder of the paper (summarized just before the beginning of
Sec.\ \ref{sec:HomogeneousHalfSpace}) presents our detailed models for
the geological strata at the two LIGO sites, and our analyses of the
various seismic modes that those strata can support and of the seismic
gravity-gradient noise produced by each of those modes.

\subsection{Phenomenology of ambient seismic motions
in the LIGO frequency band}
\label{sec:SitePhenomenology}

Seismic motions are conventionally decomposed into two components
{\cite{landau_lifshitz,love,pilant,eringen}}: {\it P-waves} and {\it
S-waves}.  P-waves have material displacements along the propagation
direction, a restoring force due to longitudinal stress (pressure ---
hence the name P-waves), and a propagation speed determined by the
material's density $\rho$ and bulk and shear moduli $K$ and $\mu$:
\begin{equation}
c_P = \sqrt{K + 4\mu/3 \over \rho}\;.
\label{eq:cP}
\end{equation}
S-waves have transverse displacements, restoring force due to shear
stress, and propagation speed
\begin{equation}
c_S = \sqrt{\mu\over\rho} = \sqrt{1-2\nu\over 2-2\nu}\;c_P \sim
{c_P\over 2}\;.
\label{eq:cS}
\end{equation}
Here $\nu$ is the material's Poisson ratio 
\begin{equation}
\nu = {3K-2\mu \over 2(3K+\mu)}\;.
\label{eq:nu}
\end{equation}
Near the earth's surface, where seismic gravity-gradient noise is
generated, these speeds are in the range $c_P\sim 500$--$2000$ m/s and
$c_S \sim 250$--$700$ m/s.  However, some of the modes that may
contribute to the noise extend down to much greater depths, even into
the bedrock where $c_P \sim 5000$--$6000$ m/s and $c_S \sim 3200$ m/s.

The ambient seismic motions are a mixture of P-waves and S-waves that
propagate horizontally (``surface waves''), confined near the earth's
surface by horizontal geological strata.  Depending on the mode type
and frequency, the horizontal propagation speed $c_H$ can range from
the surface layers' lowest S-speed to the bedrock's highest P-speed:
$250\,{\rm m/s}\,\alt\,c_H\alt 6000\,{\rm m/s}$.

P- and S-waves are coupled by geological inhomogeneities (typically
discontinuities at geological strata) and by a boundary condition at
the earth's surface.  At both LIGO sites the strata are alluvial
deposits above bedrock, with discontinuities that are horizontal to
within $2$ degrees (more typically to within less than $1$ degree).
Throughout this paper we shall approximate the material as precisely
horizontally stratified.

Seismic gravity-gradient noise is a potentially serious issue in the
frequency band from $f\sim 3$ Hz (the lowest frequency at which
mechanical seismic isolation looks practical) to $f\sim 30$ Hz; {\it
cf.}\ Fig.\ {\ref{fig:hSGG}} below.  In this frequency band, the
wavelengths of P- and S-waves are
\begin{equation}
\lambda_P = 100\,{\rm m} {(c_P/1000\,{\rm ms}^{-1}) \over
(f/10\,{\rm Hz})}, \;
\lambda_S = 50\,{\rm m} {(c_S/500\,{\rm ms}^{-1}) \over
(f/10\,{\rm Hz})}.
\label{eq:lambda}
\end{equation}

Neglecting coupling, the amplitudes of these waves attenuate as
$\exp(-\pi r/Q\lambda)$, where $r$ is the distance the waves have
propagated and $Q$ is the waves' quality factor.  The dominant
dissipation is produced by the waves' shear motions and can be thought
of as arising from an imaginary part of the shear modulus in
expressions (\ref{eq:cP}) and (\ref{eq:cS}) for the propagation speeds
$c_S$ and $c_P$ (and thence also from an imaginary part of the
propagation speeds themselves).  Since the restoring force for S-waves
is entirely due to shear, and for P-waves only about half due to
shear, the S-waves attenuate about twice as strongly as the
P-waves. The measured $Q$-factors for near-surface materials are $Q_S
\sim 10$--$25$, $Q_P \sim 20$--$50$ {\cite{abercombie,ishihara}},
corresponding to amplitude attenuation lengths
\begin{eqnarray}
{\cal L}_P &=& {Q_P \lambda_P\over\pi} 
= 1000\,{\rm m} {(Q_P/30)(c_P/1000\,{\rm ms}^{-1}) \over
f/10\,{\rm Hz}}\;, \nonumber\\ 
{\cal L}_S &=& {Q_S \lambda_S\over\pi} 
= 250\,{\rm m} {(Q_S/15)(c_S/500\,{\rm ms}^{-1}) \over
f/10\,{\rm Hz}}\;.
\label{eq:attenuation}
\end{eqnarray}
For bedrock (and basalt that overlies it at Hanford), the $Q$'s and
attenuation lengths can be higher than this --- $Q_P$ as high as a few
hundred {\cite{CRC}}.

Shallowly seated wave modes which cause ambient seismic motions in our
band, {\it i.e.}, modes that are confined to the alluvia so $c_H \alt
2500$ m/s (and more typically $\alt 1000$ m/s), must be generated in
the vicinity of the interferometers' corner and end stations by
surface sources such as wind, rain, and human activities (automobile
traffic, sound waves from airplanes, {\it etc.}); their attenuation
lengths are too short to be generated from further than a kilometer or
so.  Deep seated modes that reach into the bedrock could originate
from rather further away --- at $10$ Hz and in a layer that has
$Q_P\sim100$, $c_P\sim 5500$ m/s, modes can propagate as far as $\sim
20$ km.

In horizontally stratified material, the wave components that make up
each mode all propagate with the same angular frequency $\omega = 2\pi
f$, horizontal wave vector $\vec k = k \hat k$ (where $\hat k$ is
their horizontal direction, and $k=2 \pi / \lambda$ their horizontal
wavenumber), and horizontal phase speed $c_H = \omega/k$.  Their
vertical motions differ from one horizontal layer to another and from
P-component to S-component.  The horizontal dispersion relation
$\omega(k)$ [or equivalently $c_H(f)$] depends on the mode (Figs.\
\ref{fig:GeomOptDRHanford}, \ref{fig:DRHanford},
\ref{fig:DRLivingston}, and \ref{fig:DRLivingstonP} below).

Geophysicists divide these surface normal modes into two types
{\cite{pilant,eringen}}:
\begin{itemize}
\item
{\it Love modes,} which we shall denote by L.  These are S-waves with
horizontal displacements (``SH-waves'') that resonate in the
near-surface strata.  They involve no P-waves and thus have no
compression and no density variations; therefore, they produce no
fluctuating gravitational fields and no seismic gravity-gradient
noise.
\item
{\it Rayleigh modes,} which we shall denote by R.  These are
combinations of S-waves with vertical displacements and P-waves
(``SV-waves'') that are coupled by the horizontal discontinuities at
strata interfaces, including the earth's surface.  Rayleigh modes are
the producers of seismic gravity-gradient noise.
\end{itemize}

We shall divide the Rayleigh modes into two groups: the {\it
fundamental} Rayleigh mode, which we denote RF, and Rayleigh {\it
overtones} (all the other modes).  Rayleigh overtones require
stratification of the geological structure in order to be present;
they essentially consist of coupled SV- and P-waves which bounce and
resonate between the earth's surface and the interfaces between
strata.  We shall further divide the Rayleigh overtones into two broad
classes: those that are composed predominantly of SV-waves, denoted
RS, and those composed predominantly of P-waves, denoted RP.  In the
geophysics literature, the modes we identify as RP are sometimes
referred to simply as P-modes, and our RS modes are referred to as the
Rayleigh overtones.  However, when RP modes are intermixed with RS
modes in the $(c_H,f)$ space of dispersion relations (as turns out to
be the case at Hanford; {\it cf.}\ Fig.\ {\ref{fig:DRHanford}} below),
a given Rayleigh overtone will continuously change character from RS
to RP.  Because this will be quite important for the details of the
seismic gravity-gradient noise, we prefer to emphasize the
similarities of the two mode types by designating them both as
Rayleigh overtones and denoting them RS and RP.

We shall append to each Rayleigh overtone an integer that identifies
its order in increasing horizontal speed $c_H$ at fixed frequency $f$.
Each successive Rayleigh mode, RF, RS1, RS2, $\ldots$ (and, as a
separate series, RF, RP1, RP2, $\ldots$) penetrates more deeply into
the earth than the previous one.  In our frequency band, the
fundamental RF is typically confined to within $\sim \lambda_S/\pi
\sim 10$ m of the earth's surface.

The RF mode is evanescent in all layers (except, at low frequencies,
in the top layer).  The overtones RS1, RS2, $\ldots$ are composed
primarily of SV-waves that propagate downward from the earth's
surface, reflect off some interface, return to the surface and reflect
back downward in phase with the original downward propagating waves,
thereby guaranteeing resonance.  On each reflection and at each
interface between layers, these modes generate a non-negligible
admixture of P-waves.  The RP overtones are similar to RS, but with
the propagating and reflecting waves being largely P with some
non-negligible accompanying SV.

Dissipation will cause an overtone's waves to damp out with depth.  If
that damping is substantial in traveling from the surface to the
reflection point, the overtone will not resonate and will be hard to
excite.  Roughly speaking, the amount of amplitude decay in traveling
from the surface to the reflection point and back to the surface is $n
\pi /Q$ where $n$ is the mode number (or equivalently the number of
round-trip wavelengths); {\it cf.}\ Eqs.\ (\ref{eq:attenuation}). The
round-trip damping therefore exceeds $1/e$ for mode numbers $n \agt
Q_S/\pi\sim 5$ for RS modes and $n \agt Q_P/\pi\sim 10$ for RP modes.
Correspondingly, in this paper we shall confine attention to modes
with mode numbers $n\alt 10$.

The RP modes are harder to analyze with our formalism than RS modes
--- typically, when RP modes turn on, there are many modes very
closely spaced together and it is difficult to distinguish them.  For
this reason, we shall study only the lowest one at each site, RP1,
plus RP modes that travel nearly horizontally in the several-km thick
basalt layer at Hanford.  We expect RP1 to be typical of other
low-order RP modes, and the basalt-layer RP waves to be typical also
of such waves propagating nearly horizontally in the bedrock.

\subsection{Transfer functions and anisotropy ratio}
\label{sec:Tfunction}

Following Saulson \cite{saulson_monograph}, we shall embody the
results of our gravity-gradient analysis in a {\it transfer function}
\begin{equation}
T(f) \equiv {\tilde x(f)\over \tilde W(f)}
\label{eq:Tdef}
\end{equation}
from seismic-induced earth motions $\tilde W(f)$ to differential
test-mass motion $\tilde x(f)$.  The precise definitions of $\tilde
W(f)$ and $\tilde x(f)$ are as follows:

We shall denote the square root of the spectral density (the
``spectrum'') of the earth's horizontal surface displacements along
some arbitrary horizontal direction by $\tilde X(f)$ (units ${\rm
m}/\sqrt{\rm Hz}$), where $f$ is frequency.  We assume that $\tilde
X(f)$ is independent of the chosen direction, {\it i.e.}\ the seismic
motions are horizontally isotropic.  This is justified by seismometer
measurements at the LIGO sites before construction began
{\cite{rohaylston,rohayhanf}} and by rough estimates of the
diffractive influence of the constructed facilities (Sec.\
{\ref{sec:Conclusions}}).  We shall denote the spectrum of vertical
displacements at the earth's surface by $\tilde Z(f)$.  The quantity
$\tilde W(f)$ that appears in the transfer function is the
displacement rms-averaged over 3-dimensional directions:
\begin{equation}
\tilde W(f) = \sqrt{2\tilde X^2(f) + \tilde Z^2(f) \over 3}\;.
\label{eq:W}
\end{equation}

The other quantity, $\tilde x(f)$, which appears in the transfer
function (\ref{eq:Tdef}) is related to the interferometer's
gravitational-wave strain noise spectrum $\tilde h(f)$ by $\tilde x(f)
\equiv \tilde h(f) L$, where $L$ is the interferometer arm length (4
km for LIGO).  Physically, $\tilde x(f)$ is the spectrum of the
interferometer's arm-length difference and is called the
interferometer's ``displacement noise spectrum''.  Since $\tilde x(f)$
and $\tilde W(f)$ both have units of ${\rm m}/\sqrt{\rm Hz}$, the
transfer function $T(f)$ is dimensionless.

In this paper we shall express $T(f)$ in terms of a dimensionless
correction $\beta(f)$ to a simple and elegant formula that Saulson
\cite{saulson83} derived:
\begin{eqnarray}
T(f) \equiv {\tilde x(f) \over \tilde W(f)} &=& {4\pi
G\rho\over\sqrt{(\omega^2 - \omega_0^2)^2 + {\omega}^2/\tau^2}}
\beta(f)\nonumber\\
&\simeq& {4\pi G\rho\over(2\pi f)^2}\beta(f) \quad \mbox{at }f\agt
3\,{\rm Hz}\;.
\label{eq:Tbeta}
\end{eqnarray}
Here $\rho\simeq 1.8\,{\rm g/cm}^3$ is the mass density of the earth
in the vicinity of the interferometer, $G$ is Newton's gravitational
constant, $\omega = 2\pi f$ is the angular frequency of the seismic
waves and their fluctuating gravitational forces, and $\omega_0 \sim
2\pi$ rad/s and $\tau \sim 10^8\,{\rm s}$ are the angular frequency
and damping time of the test mass's pendular swing. We shall call
$\beta(f)$ the {\it reduced transfer function}.  Saulson's estimate
for $\beta(f)$ was
\begin{equation}
\beta_{\rm Saulson} = 1/\sqrt3 = 0.58\;;
\label{betaSaulson}
\end{equation}
{\it cf.}\ Eq.\ (21) of Ref.\ \cite{saulson83}.  Our analyses
(below) suggest that at quiet times $\beta$ may be 
$\simeq 0.35$ to 0.6, and at noisy times, $\beta\simeq 0.15$ to 1.4.
Thus, Saulson's rough estimate was remarkably good. 

Each mode of the earth's motion will contribute to the transfer
function, and since the relative phases of the modes should be
uncorrelated, they will contribute to $\beta(f)$ in quadrature:
\begin{equation}
\beta = \sqrt{\sum_{J} w_J {\beta_J}^2}\;.
\label{eq:betamixed}
\end{equation}
The sum runs over all Rayleigh and Love modes, $J\in({\rm RF},{\rm
RS}n,{\rm RP}n,{\rm L}n)$; $\beta_J(f)$ is the reduced transfer
function for mode $J$, with
\begin{equation}
\beta_{Ln} = 0
\end{equation}
because the Love modes produce no gravity-gradient noise.  The
weighting factor $w_J$ is the fractional contribution of mode $J$ to
the mean square seismic displacement $\tilde W^2$, and correspondingly
the $w_J$'s are normalized by
\begin{equation}
\sum_J w_J = 1\;.
\label{eq:sumaJ}
\end{equation}

Besides this normalization condition, there is another constraint on
the weighting factors $w_J$: each mode (at each frequency) has its own
ratio ${\cal A}_J$ of vertical to horizontal displacement at the
earth's surface:
\begin{equation}
{\cal A}_J(f) = {\tilde Z_J(f)\over \tilde X_J(f)}\;.
\label{eq:calAJ}
\end{equation}
We shall call this ratio the mode's {\it anisotropy
ratio}\footnote{Geophysicists use the name {\it spectral ratio} for
$1/{\cal A} = 1/($anisotropy ratio).}.  Since the Love modes have
purely horizontal motions, their anisotropy ratios vanish:
\begin{equation}
{\cal A}_{Ln} = 0\;.
\end{equation}
It is straightforward to show that the anisotropy ratios for the
various modes combine to produce the following net anisotropy in the
earth's surface displacement:
\begin{equation}
{\cal A}\equiv {\tilde Z\over\tilde X} 
= {\sqrt{\sum_J w_J{\cal A}_J^2/(2+{\cal A}_J^2)}\over 
\sqrt{\sum_J w_J/(2+{\cal A}_J^2)} }\;.
\label{eq:calA}
\end{equation}
At quiet times, measurements show this to be near unity at Hanford
{\cite{rohayhanf}}, and $\sim 0.6$ at Livingston {\cite{rohaylston}},
while at noisy times it can fluctuate from $\sim 0.2$ to $\sim 5$.
The measured value of this ratio is an important constraint on the
mixture of modes that produces the observed seismic noise and thence
on the net reduced transfer function.  For example, if the observed
noise is due to one specific Rayleigh mode $J$ with large anisotropy
ratio ${\cal A}_J$, accompanied by enough Love waves to reduce the net
anisotropy ratio to ${\cal A}_{\rm net} = 1.0$ (Hanford) or 0.6
(Livingston), then Eqs.\ (\ref{eq:betamixed})--(\ref{eq:calA}) imply
that the net reduced transfer function for the seismic gravity
gradient noise is
\begin{equation}
\beta_{J L} = \beta_J \sqrt{1+2/{\cal A}_J^2 \over 1+2/{\cal A}_{\rm net}^2}\;.
\label{eq:betaJL}
\end{equation}

\begin{figure}
\epsfxsize=2.8in\epsfbox{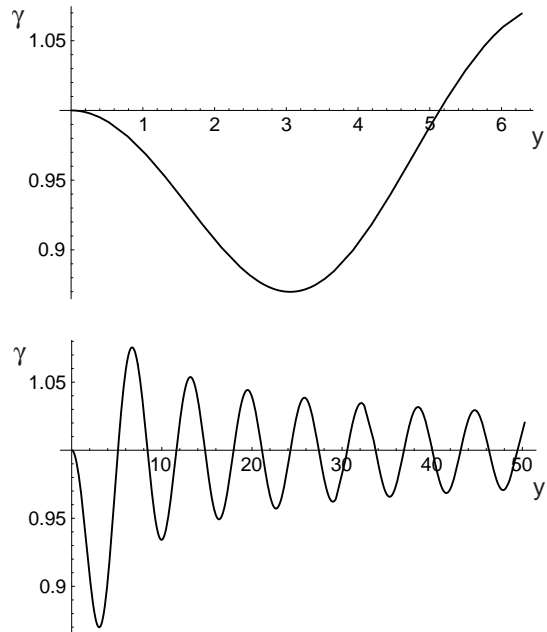}
\caption{The function $\gamma(y)$ that 
accounts for correlations of seismic gravity-gradient noise in the two
corner test masses.  This function is given analytically by Eq.\
(\protect\ref{eq:gamma}), and it appears in all of the reduced
transfer functions: $\beta_J(f)=\beta'_J(f) \gamma(2\pi f l/c_{H
J})\Gamma_J(f)$. }
\label{fig:gamma}
\end{figure}

In Appendix \ref{app:betap} it is shown that for each mode $J$, the
reduced transfer function $\beta_J$ can be split into the product of
three terms:
\begin{equation}
\beta_J = \gamma_J \Gamma_J \beta'_J\;.
\label{eq:betaJsplit}
\end{equation}
The first term, $\gamma_J$, accounts for the correlation between the
gravity-gradient noise at the interferometer's two corner test masses.
It is a universal, mode-independent function of the waves' horizontal
phase shift in traveling from one test mass to the other:
\begin{equation}
\gamma_J = \gamma(\omega l/c_{H J})\;.
\label{eq:gammaJ}
\end{equation}
Here $\omega = 2\pi f$ is the waves' angular frequency, $l\sim 5$ m is
the distance between the two corner test masses, $c_{H J}$ is the
horizontal phase speed $c_H$ for mode $J$, and $\omega/c_H \equiv k$
is the mode's horizontal wave number.  For frequencies and modes of
interest to us, the argument $y=\omega l/c_{H J}$ of $\gamma$ is of
order unity.  The function $\gamma(y)$, given by
\begin{equation}
\gamma(y) \equiv \sqrt{1+ {1\over2\pi} \int_0^{2\pi} \!\!\!\!\cos\phi \sin\phi
\cos\left(y {\cos\phi + \sin\phi \over \sqrt2} \right)d\phi}\;,
\label{eq:gamma}
\end{equation}
is plotted in Fig.\
{\ref{fig:gamma}}. As
Fig.\ {\ref{fig:gamma}} shows, $\gamma(y)$ is within about 10 per cent
of unity for all frequencies, so we shall regard it as unity througout
the rest of this manuscript, except in Appendix  \ref{app:betap}.  

The second term, $\Gamma_J$, in Eq.\ (\ref{eq:betaJsplit}) for
$\beta_J$ describes the attenuation of the gravity gradient noise due
to the height $\cal H$ of the test masses above the earth's surface.
We show in Appendix
\ref{app:betap} that 
\begin{equation}
\Gamma_J = \exp({-\omega {\cal H} /c_{H J}})\;.
\label{eq:GammaJ}
\end{equation}
For LIGO interferometers $\cal H$ is about 1.5 m, the frequency of
greatest concern is $f = \omega/2\pi \simeq 10$ Hz ({\it cf.}\ Fig.\
\ref{fig:hSGG} below), and at quiet times the dominant contribution to
the noise probably comes from the RF mode ({\it cf.}\ Sec.\
\ref{subsec:OurModes}) for which, near 10 Hz, $c_H \simeq 330$ m/s
({\it cf.}\ Figs.\ {\ref{fig:DRHanford}} and \ref{fig:DRLivingston});
correspondingly, $\Gamma_{RF}\simeq 0.75$.  For other modes, $c_H$
will be larger so $\Gamma_J$ will be closer to unity than this. For
this reason, throughout the rest of this paper, except in Appendix
{\ref{app:betap}}, we shall approximate $\Gamma_J$ by unity.  With
$\gamma_J$ and $\Gamma_J$ both approximated as unity, we henceforth
shall blur the distinction between $\beta_J$ and $\beta'_J$, treating
them as equal [{\it cf.}\ Eq.\ (\ref{eq:betaJsplit})].

In Appendix \ref{app:betap} we derive expressions for the reduced
transfer function $\beta'_J(f)$ and the anisotropy ratio ${\cal A}_J$
in terms of properties of the eigenfunctions for mode $J$: denote by
$\xi_{H J}$ and $\xi_{V J}$ the mode's complex amplitudes at the
earth's surface ($z=0)$ for horizontal displacement and {\it upward}
vertical displacement, so the mode's surface displacement
eigenfunction is
\begin{equation}
\vec \xi_J  = (\xi_{H J} \hat k - \xi_{V
J} \vec e_z) e^{i(\vec k \cdot \vec x - \omega t)}\;,
\label{eq:eigenfunction}
\end{equation}
where $\vec e_z$ is the unit vector pointing {\it downward} and $\hat
k = \vec k / k$ is the unit vector along the propagation direction.
Also, denote by ${\cal R}_J(z)$ the mode's amplitude for the
fractional perturbation of density $\delta\rho/\rho$ at depth $z$
below the surface, so
\begin{equation}
{\delta \rho_J\over\rho} = \left[\xi_{V J} 
\delta(z) + {\cal R}_J(z)\right] e^{i(\vec k \cdot \vec x 
-\omega t)}\;.
\label{eq:drhoOverrho}
\end{equation}
Here the term $\xi_{V J} \delta(z)$ accounts for the mass moved above
$z=0$ by the upward vertical displacement $\xi_V$.  Then, we show in
Appendix \ref{app:betap} [Eq.\ (\ref{eq:alpha1})] that
\begin{equation}
{\cal A}_J = \sqrt2 {|\xi_{V J} | \over |\xi_{H J}|}\;,
\label{eq:calAJ1}
\end{equation}
where the $\sqrt2$ comes from the fact that when this mode is
incoherently excited over all horizontal directions $\hat k$, its rms
horizontal amplitude along any chosen direction is $|\xi_{H
J}|/\sqrt2$.  Similarly, we show in Appendix \ref{app:betap} [Eq.\
(\ref{eq:betap1})] that
\begin{equation}
\beta'_J(f) = \sqrt{3/2\over |\xi_{H J}|^2 + |\xi_{V J}|^2}\;
\left| \xi_{V J} + \int_0^\infty {\cal R}_J(z)
e^{-kz} dz\right|\;,
\label{eq:betapJ}
\end{equation}
where $k=\omega/c_{H J}$ for mode $J$.  We shall refer to the $\xi_{V
J}$ term in Eq.\ (\ref{eq:betapJ}) as the {\it surface source} of
gravity gradients, and the $\int {\cal R}_J e^{-kz} dz$ term as the
{\it subsurface source}.

Note that the influence of a given density perturbation dies out as
$e^{-kz}$, so unless ${\cal R}_J(z)$ increases significantly with
depth, the seismic gravity gradients arise largely from depths
shallower than the {\it gravity-gradient e-folding length}
\begin{equation}
{\cal Z}_{\rm sgg} = {1\over k} = {c_{H J}
\over 2\pi f} = 16 {\rm m} {(c_{H J}/1000
{\rm m} {\rm s}^{-1}) \over (f/10 \rm{Hz})}\;.
\label{eq:sggdepth}
\end{equation}
This has a simple explanation: (i) to produce much gravitational force
on a test mass, a compressed bit of matter must reside at an angle
$\alpha \agt \pi/4$ to the vertical as seen by the test mass, and (ii)
bits of matter all at the same $\alpha \agt \pi/4$ and at fixed time
have fractional compressions $\delta\rho/\rho$ that oscillate with
depth $z$ as $e^{ikx} = e^{ikz\tan\alpha}$, and that therefore tend to
cancel each other out below a depth $1/(k\tan\alpha) \sim 1/k$.

From Eq.\ (\ref{eq:betapJ}) we can estimate the magnitude of the
reduced transfer function.  The mode's fractional density perturbation
${\cal R}_J$ is equal to the divergence of its displacement
eigenfuction (aside from sign), which is roughly $k \xi_{H J}$ and
often does not vary substantially over the shallow depths $z \alt
{\cal Z}_{\rm sgg}$ where the gravity gradients originate.
Correspondingly, the integral in Eq.\ (\ref{eq:betapJ}) is $\sim\xi_{H
J}$, so $\beta'_J\sim \sqrt{1.5 |\xi_{H J} + \xi_{V J}|^2/(|\xi_{H
J}|^2 + |\xi_{V J}|^2)}\sim 1$, since the horizontal and vertical
displacements are comparable.

As we shall see in Secs.\ {\ref{sec:HomogeneousHalfSpace}} and
{\ref{subsec:AbetapHanford}} below, for RP modes the gravity gradients
produced by the surface and subsurface sources tend to cancel (a
consequence of mass conservation), so $\beta'$ actually tends to be
somewhat smaller than unity,
\begin{mathletters}
\label{eq:betapRPRFRS}
\begin{equation}
\beta'_{\rm RP} \alt 0.15\;,
\label{eq:betapRP}
\end{equation}
while for RF and RS modes, the surface source tends to dominate, so  
\begin{equation}
\beta'_{\rm RF} \sim \beta'_{\rm RS} \sim
{1\over\sqrt2} \sqrt{3\over 1+2{\cal A}_J^2} \sim {1\over\sqrt2}
=0.7\;.
\label{eq:betapRFRS}
\end{equation}
\end{mathletters}
If we had normalized our transfer function to the vertical
displacement spectrum $|\tilde Z(f)|$ instead of the
direction-averaged spectrum $|\tilde W(f)|$ [Eq.\ (\ref{eq:Tdef})],
then for modes in which the surface source strongly dominates,
$\beta'_J$ would be $1/\sqrt2$ independently of the mode's anisotropy
ratio.

In Secs.\ \ref{sec:Hanford} and \ref{sec:Livingston} and associated
Appendices, we shall derive, for each low-order Rayleigh mode at
Hanford and Livingston, the reduced transfer function $\beta'_J$ and
the anisotropy ratio ${\cal A}_J$.  In Sec.\ \ref{sec:Tus}, we shall
discuss the likely and the allowed weightings $w_J$ of the various
modes [subject to the constraints (\ref{eq:sumaJ}) and
(\ref{eq:calA})], and shall estimate the resulting net reduced
transfer functions $\beta(f)$ for the two sites and for quiet and
noisy times (Table \ref{table:betap}).

Henceforth we typically shall omit the subscript $J$ that denotes the
mode name, except where it is needed for clarity.

\subsection{Saulson's analysis and transfer function}
\label{sec:Tsaulson}

In his original 1983 analysis of seismic gravity-gradient noise
{\cite{saulson83}}, Saulson was only seeking a first rough estimate,
so he used a fairly crude model.  He divided the earth near a test
mass into regions with size $\lambda_P /2$ (where $\lambda_P$ is the
wavelength of a seismic P-wave), and he idealized the masses of these
regions as fluctuating randomly and independently of each other due to
an isotropic distribution of passing P-waves.  Saulson's final
analytic result [his Eq.\ (21)] was the transfer function
(\ref{eq:Tdef}) with $\beta=1/\sqrt3$.

Saulson's 1983 numerical estimates\cite{saulson83} of the seismic
gravity-gradient noise were based on seismic noise levels $\tilde W(f)
= 0.5 \times 10^{-8} (10\,{\rm Hz}/f)^2 {\rm cm/\sqrt{Hz}}$ for
``average sites'' and a factor 10 lower than this for ``quiet sites''.
The resulting gravity-gradient noise $\tilde x(f) = T(f)\tilde W(f)$
was substantially below the projected vibrational seismic noise in
(seismically well isolated) ``advanced'' LIGO interferometers
{\cite{abramovici}}.

In updating these estimates for his recent monograph
{\cite{saulson_monograph}}, Saulson noted that his original
``average'' and ``quiet'' sites were based on measurements at
underground seismological stations.  Surface sites, such as those
chosen for LIGO and VIRGO, are far noisier than underground sites in
the relevant frequency band, $3 \mbox{\,Hz} \alt f \alt 30
{\mbox{\,Hz}}$, because of surface seismic waves.  More specifically,
even though the chosen LIGO sites (at Hanford, Washington and
Livingston, Louisiana) are among the more quiet locations that were
studied in the LIGO site survey, their noise at typical times is
approximately isotropic [$\tilde Z(f) \sim\tilde X(f)\sim\tilde W(f)$]
and has approximately the following form and magnitude
{\cite{rohaylston,rohayhanf}}
\begin{eqnarray}
\tilde W(f) &=& 1\times10^{-7}{\rm cm\over\sqrt{Hz}}
\quad \mbox{at }1<f<10\,{\rm Hz}\;, \nonumber\\
&=& 1\times10^{-7}{\rm cm\over\sqrt{Hz}}\left({\rm 10 Hz}\over
f\right)^2 \quad \mbox{at }f>10\,{\rm Hz}.
\label{eq:ligo_seismic_spectrum}
\end{eqnarray}
This so-called {\it standard LIGO seismic spectrum} is 20 times larger
than at Saulson's original ``average'' sites for $f\ge 10$ Hz.
Correspondingly, Saulson pointed out in his update, the seismic
gravity-gradient noise may stick up above the vibrational seismic
noise in ``advanced'' LIGO interferometers.\footnote{Saulson informs
us that in evaluating the noise at the LIGO sites, he made an error of
$\sqrt3$; his transfer function and the standard LIGO seismic spectrum
actually predict a noise level $\sqrt3$ smaller than he shows in Fig.\
8.7 of his book {\cite{saulson_monograph}}.  When this is corrected,
his predicted noise, like ours, is below the ``advanced'' LIGO noise
curve, though only slightly so near 10 Hz.}  On the other hand, at
very quiet times --- at night and with winds below 5 mph --- the LIGO
seismic ground noise $\tilde W(f)$ can be as low as $\sim 1/10$ the
level (\ref{eq:ligo_seismic_spectrum}), thereby pushing Saulson's
seismic gravity-gradient noise well below the vibrational seismic
noise of an ``advanced'' LIGO interferometer.

\subsection{Our analysis and transfer function}
\label{sec:Tus}

Saulson's new, more pessimistic estimates of the seismic gravity
gradient noise triggered us to revisit his derivation of the transfer
function $T(f)$ from seismic ground motions to detector noise.  Our
analysis consists of:

(i) splitting the ambient seismic motions into Love and Rayleigh modes
(body of this paper and appendices);

(ii) computing the reduced transfer function for each mode and for
models of the geological strata at each LIGO site (body and
appendices);

(iii) using seismic measurements at the LIGO sites and geophysical
lore based on other sites to estimate the mode mixture present at the
two sites under both quiet and noisy conditions (this section); and

(iv) evaluating for these mode mixtures the expected reduced transfer
function and resulting noise (this section).

\subsubsection{Our reduced transfer functions}
\label{subsec:OurBetap}

Table {\ref{table:betap}} summarizes the results of our model
computations for each LIGO site.  Shown there are the range of
computed reduced transfer functions $\beta'$ for specific types of
Rayleigh modes, and the range of net reduced transfer functions
$\beta'_L$ that would result if each Rayleigh mode were mixed with
enough Love waves to bring its (often rather high) anisotropy ratio
$\cal A$ down to the level typical of quiet times at the LIGO sites
(${\cal A}\simeq 1.0$ at Hanford {\cite{rohayhanf}}, ${\cal A}\simeq
0.6$ at Livingston {\cite{rohaylston}}).

\begin{table}
\caption{Reduced transfer functions $\beta'$ predicted for Hanford and
Livingston by our 4-layer models; and $\beta'_L,$ the value of
$\beta'$ when enough Love waves are added to bring the anisotropy
ratio down to the quiet-time values observed at the two sites (${\cal
A} \simeq 1$ for Hanford, ${\cal A} \simeq 0.6$ for Livingston).
\label{table:betap}
}
\begin{tabular}{lllll}
Modes&Hanford&Hanford&Livingston&Livingston\\
&$\beta'$&$\beta'_L$&$\beta'$&$\beta'_L$\\
\tableline
RF $f<10$Hz&0.4--0.85&0.35--0.6&0.65--0.9 &0.35--0.45 \\
RF $f>10$Hz&0.85&0.6&0.65--0.9&0.35--0.45 \\
RS &0.4--1.4&0.4--1.05&0--1.2 &0--0.9 \\ 
RP &0--0.15&0--0.15&0.02--0.13&0.01--0.06 \\
\end{tabular}
\end{table}

The modes shown in Table {\ref{table:betap}} are the RF mode, the RS
modes with no sign of RP admixture, and the RP modes.  The RF and RS
modes usually have $\beta'$ in the range 0.4 to 1.2, though in special
cases it can sink toward zero.  By contrast, the RP modes always have
small $\beta'$: 0 to 0.15.  This marked difference arises from the
fact that for RF and RS the (largely S-wave) surface source tends to
dominate over the (entirely P-wave) subsurface source; while for RP,
mass conservation guarantees that the two sources (both largely
P-wave) will be nearly equal, but opposite in sign, and will nearly
cancel.  (If the surface source were absent, the pattern would be
reversed: the subsurface source $\int {\cal R}$, which arises from
compressional density perturbations, tends to be weak for RS modes
because they consist primarily of non-compressional S-waves, but is
strong for RP modes since they consist primarily of compressional
P-waves; so $\beta'$ would be small for RS and large for RP.)

\subsubsection{Modes actually present and resulting seismic noise}
\label{subsec:OurModes}

There is little {\it direct} evidence regarding which modes contribute
to the ambient surface motions and thence to the gravity-gradient
noise at the LIGO sites during quiet times.  Past seismic measurements
do not shed much light on this issue.  In the concluding section of
this paper (Sec.\ \ref{sec:Conclusions}), we shall propose
measurements that could do so.

Fortunately, the nature of the ambiently excited modes has been
studied at other, geophysically similar sites (horizontally stratified
alluvia over bedrock).  The preponderance of evidence suggests that at
quiet times the surface motions at such sites and in our frequency
band are due to a mixture of Love waves and the fundamental Rayleigh
mode RF plus perhaps a few low order RS modes
{\cite{asten,douze_laster,liaw_mcevilly,powell,hough}}.  In at least
one case, some amount of RP excitation is also seen {\cite{milana}};
these RP excitations are ascribed to ``cultural noise'' (noise
generated by human activity of some sort) near the measurement site.
Deep borehole measurements indicate that RP dominates at very great
depths ($\sim 5$ km) {\cite{douze}}; this is probably not relevant to
our analysis, however.  It merely indicates that very deep down, the
majority of the surface waves have damped away, leaving only some
residual RP modes.  The deep motions are typically an order of
magnitude or two smaller than the surface motions; {\em cf.} Sec.\
\ref{sec:Tsaulson}.  

On this basis, we presume that at quiet times the net reduced transfer
function is about that for the RF mode, with enough admixed Love waves
to bring the net $\cal A$ down to the typical quiet-time values of 1.0
for Hanford and 0.6 for Livingston.  In other words, $\beta'_{\rm
net}$ is about equal to $\beta'_L$ for the RF mode:
\begin{eqnarray}
\beta'_{\rm net,\ quiet\ times} &\simeq& \mbox{0.35--0.45 at Livingston,}
\nonumber\\
&\simeq&\mbox{0.35--0.6 at Hanford.}
\label{eq:betapBestGuess}
\end{eqnarray}

\begin{figure}
\epsfxsize=3.3in\epsfbox{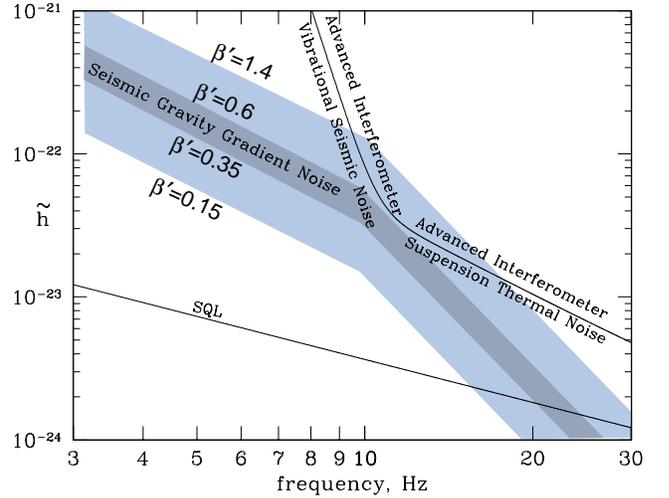}
\caption{Seismic gravity-gradient noise in a LIGO interferometer.
In this figure, we assume that the direction-averaged spectrum of
earth displacements has the form of the standard LIGO seismic
spectrum, Eq.\ ({\protect\ref{eq:ligo_seismic_spectrum}}).  The edges
of the gray bands are for the indicated values of the reduced transfer
function $\beta'$ (assumed equal to $\beta$; {\it i.e.}, for $\gamma$
and $\Gamma$ approximated as unity).  The dark gray band is our
estimate of the range of noise for quiet times.  The gray bands, both
light and dark, are for noisy times, assuming the standard LIGO
seismic spectrum (\protect\ref{eq:ligo_seismic_spectrum}). At very
quiet times, the ground spectrum can be a factor $\sim 10$ smaller
than ({\protect\ref{eq:ligo_seismic_spectrum}}), which will lower
these bands accordingly.  Conversely, at noisy times the ground
spectrum can be larger, raising these bands.  Also shown for
comparison is the projected noise in an ``advanced'' LIGO
interferometer, and the standard quantum limit (SQL) for an
interferometer with one tonne test masses.  The SQL is the square root
of Eq.\ (122) of Ref.\ {\protect\cite{300yrs}}.  The ``advanced''
interferometer noise is taken from Fig.\ 7 of Ref.\
{\protect\cite{abramovici}}, with correction of a factor 3 error in
the suspension thermal segment (Fig.\ 7 of Ref.\
{\protect\cite{abramovici}} is a factor 3 too small, but Fig.\ 10 of
that reference is correct, for the parameters listed at the end of the
section ``LIGO Interferometers and Their Noise'').}
\label{fig:hSGG}
\end{figure}

We have folded these quiet-time estimates for $\beta'$ into the
standard LIGO seismic spectrum (\ref{eq:ligo_seismic_spectrum}) to
obtain the gravity-gradient noise estimates shown as the dark gray
band in Fig.\ {\ref{fig:hSGG}}.  The thickness of the band indicates
the range of our $\beta'$ [Eq.\ ({\ref{eq:betapBestGuess}})]: $0.35$
to $0.6$.  To produce this plot, we took expression (\ref{eq:Tbeta})
for the transfer function $T(f)$ with $\gamma$ and $\Gamma$ set to
unity, so $\beta=\beta'$ [{\it cf.}\ Eq.\ (\ref{eq:betaJsplit})].
Then, we multiplied this by the standard LIGO seismic spectrum
(\ref{eq:ligo_seismic_spectrum}) for the ground displacement with an
assumed density $\rho = 1.8$ g/cm$^3$.  This yields
\begin{eqnarray}
\tilde h_{\rm SGG}(f) &=& {\beta'\over 0.6}
{6 \times 10^{-23}\over \sqrt{\rm Hz}} \left({{\rm
10 Hz}\over f}\right)^2\!\!, \; 3\,{\rm Hz} \alt f < 10\,{\rm Hz},
\nonumber\\
&=& {\beta'\over 0.6} {6 \times 10^{-23} \over\sqrt{\rm Hz}} \left({{\rm
10 Hz}\over f}\right)^4\!\!, \; 10\,{\rm Hz} < f \alt 30\,{\rm Hz},
\nonumber\\
\label{eq:hSGG}
\end{eqnarray}
which we plotted for the indicated values of $\beta'$.

At very quiet times, the ambient seismic spectrum near 10 Hz can be as
much as a factor $\sim 10$ lower than the standard LIGO spectrum
assumed in Eq.\ (\ref{eq:hSGG}) and Fig.\ \ref{fig:hSGG}, and
correspondingly the quiet-time gravity gradient noise can be a factor
$\sim 10$ lower.

At noisier times, there appear to be excitations of a variety of RF,
RS and RP modes.  For example, at the LIGO sites, time delays in
correlations between surface motions at the corner and the end
stations reveal horizontal propagation speeds $c_H \sim 5000$ m/s,
corresponding to deeply seated RP-modes (although for the most part
these modes are seen at frequencies too low to be of interest in this
analysis --- $f\alt 0.2$ Hz {\cite{rohaydiff,rohayemail}}).  Moreover,
the measured anisotropy ratios can fluctuate wildly from $\sim 0.2$ to
$\sim 5$ at noisy times, suggesting a wildly fluctuating mixture of
RF, RS, RP, and Love modes.  Scrutinizing not only Table
{\ref{table:betap}} but also the range of $\beta'$ shown in Figs.\
{\ref{fig:PropertiesHanford}}, {\ref{fig:RPHHanford}},
{\ref{fig:PropertiesLivingston}} and {\ref{fig:DRLivingstonP}} which
underlie that table, and keeping in mind that Love modes with
vanishing $\beta'$ will also be present, we estimate that the
fluctuations of $\beta'$ at noisy times will be confined to the range
\begin{equation}
\beta'_{\rm net,\ noisy\ times} \simeq \mbox{0.15--1.4.}
\label{eq:betapNoisy}
\end{equation}
We have folded this estimate into the standard LIGO seismic spectrum
to obtain the upper and lower edges of the light gray band in Fig.\
{\ref{fig:hSGG}}.  The grey bands, light and grey taken together, are
our best estimate of the range of seismic gravity-gradient noise at
noisy times, assuming the standard LIGO seismic spectrum.  Since, at
noisy times, the seismic spectrum can be somewhat higher than the
standard one, the gravity-gradient noise will be correspondingly
higher.

For the next few years, the most important application of these
estimates is as a guide for the development of seismic isolation
systems and suspension systems for LIGO.  There is not much point in
pushing such systems so hard that the vibrational seismic or the
suspension thermal noise is driven far below our lowest estimates of
the seismic gravity-gradient noise [bottom of the black line in Fig.\
{\ref{fig:hSGG}}, lowered by the amount that the actual very quiet
time spectrum falls below the standard LIGO spectrum
(\ref{eq:ligo_seismic_spectrum})]---unless corresponding steps are
taken to mitigate the seismic gravity gradient noise; see Sec.\
\ref{sec:Conclusions}.

In Fig.\ {\ref{fig:hSGG}} we compare our predicted seismic gravity
gradient noise to the projected noise in ``advanced'' LIGO
interferometers and to the standard quantum limit for an
interferometer with one tonne test masses (``SQL'').  Notice that our
lower bound on the seismic gravity-gradient noise is everywhere
smaller than the ``advanced'' interferometer noise, but it is larger
than the SQL at frequencies below $\sim 20$ Hz.  Our lower bound rises
large enough below $\sim 10$ Hz to place limits on seismic-isolation
and suspension-noise R\&D that one might contemplate doing at such
frequencies.

The remainder of this paper is organized as follows: we begin in Sec.\
{\ref{sec:HomogeneousHalfSpace}} by discussing Rayleigh waves and
seismic gravity-gradient noise in the idealized case of a homogeneous
half space (not a bad idealization for some regions of some modes at
Hanford and Livingston).  Then we develop multilayer geophysical
models for Hanford and Livingston and use them to derive the reduced
transfer functions for the various Rayleigh modes (Secs.\
{\ref{sec:Hanford}} and {\ref{sec:Livingston}}).  We conclude in Sec.\
{\ref{sec:Conclusions}} with a discussion of the uncertainties in our
analysis and research that could be undertaken to reduce the
uncertainties, and also a discussion of the physical interaction of
the seismic waves with the foundations of the LIGO facilities, and of
ways to somewhat reduce the gravity gradient noise if it ever becomes
a serious problem in LIGO interferometers.  Mathematical details of
our analysis are confined to Appendices.  Those Appendices may form a
useful foundation for analyses of seismic gravity-gradient noise at
other sites.

\section{Homogeneous half space}
\label{sec:HomogeneousHalfSpace}

\subsection{Fundamental Rayleigh mode}
\label{subsec:HHRayleigh}

As a first rough guide to seismic gravity-gradient noise, we idealize
the LIGO sites as a homogeneous half space with density $\rho$,
Poisson ratio $\nu$, S-wave speed $c_S$ and P-wave speed $c_P$ given
by
\begin{equation}
\rho=1.8\,{\rm g/cm}^3, \;\;\nu = 0.33, \;\; c_P = 440\,{\rm m/s},
\;\; c_S = 220\,{\rm m/s}.
\label{eq:HalfSpaceModel}
\end{equation}
(These are the measured parameters of the surface material at
Livingston; for Hanford, the parameters are only a little different;
{\it cf.}\ Sec.\ {\ref{subsec:HanfordGeophysics}} below.)

This homogeneous half space can only support the RF mode, as mentioned
in the Introduction.  The theory of the RF mode and the seismic
gravity-gradient noise that it produces is sketched in Appendix
{\ref{app:HomogeneousHalfSpace}}.  Here we summarize the results.

The RF mode propagates with a horizontal speed $c_H$ that depends
solely on the Poisson ratio.  It is a bit slower than the speed of
S-waves, and is much slower than P-waves.  For the above parameters,
\begin{equation}
c_H = 0.93 c_S = 205\,{\rm m/s}\;;
\label{eq:crHH}
\end{equation}
{\it cf.}\ Eq.\ (\ref{eq:cR}).  Correspondingly, the waves' horizontal wave number
$k$ and horizontal reduced wavelength are
\begin{equation}
{\lambda\over2\pi} = {1\over k} = 3.3\,{\rm m}\left({10\,{\rm Hz}
\over f}\right)\;.
\label{eq:lambdaHH}
\end{equation} 

Because $c_H < c_S < c_P$, RF waves are evanescent vertically: the
P-waves die out with depth $z$ as $e^{-qkz}$, and the SV-waves as
$e^{-skz}$, where
\begin{eqnarray}
q&=&\sqrt{1-(c_H/ c_P)^2} = 0.88 \;, \nonumber\\
s&=&\sqrt{1-(c_H/ c_S)^2}=0.36\;.
\label{eq:qsHH}
\end{eqnarray} 
Thus, the vertical $e$-folding lengths for compression (which produces
seismic gravity gradients) and shear (which does not) are
\begin{eqnarray}
{\cal Z}_P &=& {1\over qk} = 3.7\,{\rm m}\left({10\,{\rm Hz} \over
f}\right)\;, \nonumber \\ 
{\cal Z}_S &=& {1\over sk} = 9.0\,{\rm m}\left({10\,{\rm Hz} \over
f}\right)\;.
\label{eq:LHH}
\end{eqnarray}

These RF waves produce substantially larger vertical motions than
horizontal at the earth's surface.  For waves that are horizontally
isotropic, the anisotropy ratio is
\begin{equation}
{\cal A} = \sqrt{2}{q(1-s^2)\over 1+s^2-2qs} = 2.2\;.
\label{eq:calAHH}
\end{equation}
This large ratio is indicative of the fact that RF waves contain a
large component of P-waves.  As mentioned in the Introduction, this is
substantially larger than the values typically observed at the LIGO
sites in the band $3\,{\rm Hz}\alt f\alt\,30\,{\rm Hz}$ --- seismic
measurements taken at those sites {\cite{rohaylston,rohayhanf}} show
that, at quiet times, ${\cal A}\simeq1.0$ at Hanford, ${\cal
A}\simeq0.6$ at Livingston.  Thus, RF waves cannot alone be
responsible for the seismic motions.  To the extent that our
homogeneous-half-space model is realistic, RF waves must be augmented
by a large amount of horizontally-polarized S-waves (``SH-waves''),
which have ${\cal A} = 0$.

RF waves produce a reduced transfer function 
\begin{equation}
\beta' = \sqrt{3(1+s^2-2q)^2\over 2(1+s^2)[(1+s^2)(1+q^2)-4qs]}
= 0.86\;.
\label{betapHH}
\end{equation}
This $\beta'$ is produced primarily by the surface source $\xi_V$ in
Eq.\ (\ref{eq:betapJ}); if there were no surface source, the
subsurface term $\int{\cal R}$ (arising solely from the P-wave
compressions) would produce the far smaller value $\beta'=0.17$.  When
the RF waves are augmented by enough Love waves to reduce the net
$\cal A$ to 1.0 (Hanford) or 0.6 (Livingston), they produce a net
reduced transfer function [Eq.\ (\ref{eq:betaJL}) with primes added to
the $\beta$'s]
\begin{equation}
\beta'_L = \mbox{0.59 (Hanford), 0.40 (Livingston).}
\label{eq:RFbetapL}
\end{equation}

As we shall see in the next two sections, the earth is strongly
stratified over the relevant vertical length scales at both Hanford
and Livingston, and this gives rise to significant differences from
the homogeneous-half-space model.  Nevertheless, as discussed in the
Introduction (Sec.\ {\ref{subsec:OurModes}}), it is likely that at
quiet times the RF mode produces the dominant gravity-gradient noise.
Stratification modifies this RF mode somewhat from the description
given here; however, as we shall see (Figs.\
{\ref{fig:PropertiesHanford}} and {\ref{fig:PropertiesLivingston}}),
these modifications typically alter its anisotropy ratio and reduced
transfer function by only a few tens of percent.  Thus, the
homogeneous-half-space model may be a reasonable indicator of seismic
gravity-gradient noise in LIGO at quiet times.

\subsection{P-up and SV-up waves}
\label{subsec:PUpSVUp}

The principal effect of stratification is to produce a rich variety of
normal-mode oscillations, in which mixtures of SV- and P-waves
resonate in leaky cavities formed by the strata.  These oscillations
are Rayleigh-mode overtones, whose (rather complex) theory is sketched
in Appendices \ref{app:Multilayer} and \ref{app:TwoLayer} and
discussed in Secs.\ \ref{sec:Hanford} and \ref{sec:Livingston}.  In
this subsection we will momentarily ignore that fact, and will seek
insight from a much simpler analysis that gives gives results which
agree approximately, and in some cases quite well, with those of the
Rayleigh-overtone theory.

If the top layer (labeled by a subscript 1) has a thickness $D_1$
larger than half a vertical wavelength of the waves' oscillations,
$D_1 > (c_{P1}/2 f) / {\sqrt{1-(c_{P1}/c_H)^2}}$ for P-waves and
similarly for S-waves [{\it cf.}\ Eq.\ (\ref{eq:kV}) below and
associated discussion], then the trapped modes can be thought of as
propagating upward through the top layer, reflecting at the earth's
surface, and then propagating back downward.  By ignoring the effects
of the interfaces below, these waves can be idealized as traveling in
a homogeneous half space.

\begin{figure}
\epsfxsize=3.3in\epsfbox{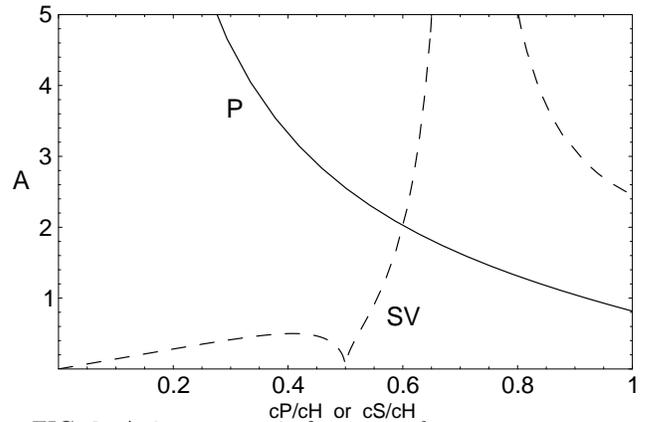}
\caption{Anisotropy ratio for waves that propagate upward in a
homogeneous half space, reflect off the Earth's surface, and propagate
back downward.  The curve ``P'' is for the case when the upward
propagating waves are pure P (P-up waves), in which case the abscissa
is $c_P/c_H\equiv\sin\alpha_P$; ``SV'' is for SV-up waves, with
abscissa $c_S/c_H\equiv\sin\alpha_S$.  It is assumed that $c_P = 2
c_S$; this is approximately the case for the surface layers at Hanford
and Livingston.}
\label{fig:AnisoUp}
\end{figure}

\begin{figure}
\epsfxsize=3.3in\epsfbox{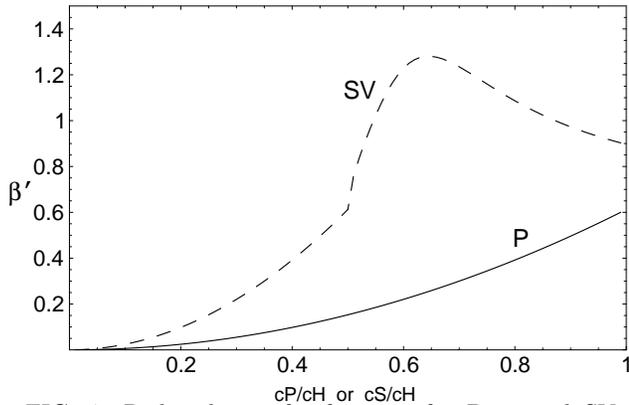}
\caption{Reduced transfer function for P-up and SV-up waves in
a homogeneous half space with $c_P = 2 c_S$.  Notation is as in Fig.\
{\protect\ref{fig:AnisoUp}}.  }
\label{fig:betapUp}
\end{figure}

The behavior of these waves depends on the mixture of P- and SV-waves
that composes them as they propagate upward.  Because these two
components will superpose linearly, we can decompose the mixture and
treat the P-wave parts and SV-wave parts separately.  We will call
these components P-up and SV-up waves.  In Appendix
{\ref{app:PupSup}}, we derive simple analytic formulae for the
anisotropy ratio $\cal A$ and reduced transfer function $\beta'$ for
P-up and SV-up waves, and in Figs.\ {\ref{fig:AnisoUp}} and
{\ref{fig:betapUp}} we graph those formulae.  In these plots, for
concreteness, we have chosen $c_S = c_P/2$.

Consider, first, the P-up waves (solid curves in Figs.\
{\ref{fig:AnisoUp}} and {\ref{fig:betapUp}}).  Due to Snell's law
[{\it cf.}\ Eq.\ (\ref{eq:Snell}) below], these waves propagate at an
angle $\alpha_P = \arcsin(c_P/c_H)$ to the vertical.  Such propagating
waves can therefore exist only for $c_H>c_P$; when $c_H < c_P$,
P-waves are evanescent.  For this reason, in the Figures we plot on
the abscissa the ratio $c_P/c_H$ running from 0 to 1.  When P-up waves
hit the surface, some of their energy is converted into SV-waves
propagating downward at an angle $\alpha_S = \arcsin(c_S/c_H)$; the
rest of the energy goes into reflected P-waves.  The resulting
combination of upgoing P- and downgoing P- and SV-waves gives rise to
the anisotropy and reduced transfer functions shown in the figures.

For $c_H\gg c_P$ the waves travel nearly vertically.  Their
P-components produce vertical motions, while the much weaker SV-waves
created on reflection produce horizontal motions.  As a result, $\cal
A$ is large, diverging in the limit $c_H \to \infty$, and decreasing
gradually to near unity as $c_H \to c_P$.  As we shall see below, this
is typical: when P-waves predominate in a wave mixture (RP modes),
$\cal A$ is typically somewhat larger than unity.

For these P-up waves, the gravity gradients produced by the surface
source cancel those from the subsurface source in the limit $c_H \gg
c_P$, causing $\beta'$ to vanish. As $c_H$ is reduced (moving
rightward in Fig.\ \ref{fig:betapUp}), the cancellation becomes
imperfect and $\beta'$ grows, though never to as large a value as
$\beta'$ would have in the absence of the surface term
($\sim1.3$--2.4).  The surface-subsurface cancellation is easily
understood.  In the limit $c_H \gg c_P$, the P-waves propagate nearly
vertically, with vertical reduced wavelength for their density
oscillations, $1/k_V = c_P/\omega$, that is small compared to the
gravity-gradient $e$-folding length ${\cal Z}_{\rm sgg} = \ 1/k =
c_H/\omega$, over which the waves' sources are integrated in Eq.\
(\ref{eq:betapJ}) to produce the gravitational force. Therefore, the
gravity gradients come from many vertical wavelengths, with adjacent
ones weighted nearly equally.  Because of mass conservation, the
surface source plus the top quarter wavelength of subsurface source
(multiplied by $\rho$) constitute the mass per unit area that has been
raised above a node of the mode's displacement eigenfunction; and
correspondingly their sum vanishes.  Below that node, alternate half
wavelengths of the subsurface source cancel each other in a manner
that gets weighted exponentially with depth, $e^{-kz}$; their
cancellation is excellent in the limit $1/k_V \ll 1/k$, {\it i.e.},
$c_H \gg c_P$.

Turn now to the SV-up waves.  Upon reflection from the surface, these
produce a mixture of downgoing SV- and P-waves.  This mixture
gives rise to the anisotropy and reduced transfer functions shown
dashed in Figs.\ {\ref{fig:AnisoUp}} and {\ref{fig:betapUp}}.  Again
by Snell's law, SV-up waves propagate at an angle $\alpha_S =
\arcsin(c_S/c_H)$ to the vertical; thus, propagation is possible only
for $c_H > c_S$, and so we plot on the abscissa $c_S/c_H$ running from
0 to 1.  When $c_H>2c_S = c_P$ (left half of graphs), the downgoing
P-waves generated at the surface can propagate; when $c_H<2c_S=c_P$
(right half of graphs), the downgoing P-waves have imaginary
propagation angle $\alpha_P$ and thus are evanescent (decay
exponentially with depth).  This is analogous to the phenomenon of
total internal reflection which one encounters in elementary optics.
The downgoing P-waves are the sole subsurface source of
gravity-gradient noise, and since they are only a modest component of
the SV-Up mode, the subsurface source is small.  The SV-waves produce
no subsurface source (no compressions), but they produce a large
surface source (large surface vertical motions).  This surface source
is the dominant cause of the gravity-gradient noise and predominantly
responsible for the rather large reduced transfer function shown in
Fig. \ref{fig:betapUp}.  Note that the maximum value, $\beta' \simeq
1.4$, is the same as the largest $\beta'$ for RS modes in our 4-layer
models of the LIGO sites (Table \ref{table:betap}).

When propagating more or less vertically ($c_H > 2c_S$), these SV-up
waves produce small anisotropies (${\cal A} < 0.4$ --- large
horizontal motions and small vertical motions).  When they propagate
more or less horizontally, $\cal A$ is large.  The divergence of $\cal
A$ at $c_S/c_H = 1/\sqrt2 = 0.707$ ($\alpha_S = \pi/4$) occurs because
the SV-up waves at this angle generate no P-waves upon reflection;
they only generate downgoing SV-waves, and the combination of the
equal-amplitude up and down SV-waves produces purely vertical motions
at the earth's surface.  At frequencies $f\agt 20$ Hz, mode RS1 at
Hanford can be approximated as an SV-up mode and exhibits this
behavior; {\it cf.}\ Sec.\ \ref{subsec:AbetapHanford}.
 
\section{Hanford}
\label{sec:Hanford}

\subsection{Hanford geophysical structure}
\label{subsec:HanfordGeophysics}

At the LIGO site near Hanford, Washington, the top 220 m consists of a
variety of alluvial layers (fluvial and glacio-fluvial deposits of the
Pliocene, Pleistocene, and Holocene eras; coarse sands and gravels,
fine sands, silts, and clays, in a variety of orders).  The upper 40 m
are dry; below about 40 m the alluvium is water-saturated.  From the
base of the alluvium (220 m) to a depth of $\sim4$ km lies a sequence
of Columbia River basalts, and below that, bedrock
{\cite{dames_moore,skagit}}.

The density of the alluvial material is $\rho\simeq 1.8$ g/cm$^3$,
independent of layer.  Velocity profiles ($c_P$ and $c_S$ as functions
of depth $z$) have been measured at the site by contractors in
connection with two projects: LIGO {\cite{dames_moore}} and the Skagit
nuclear power plant {\cite{skagit}} (which was never constructed).  We
have relied primarily on the Skagit report because it contains more
detailed information over the range of depths of concern to us, and
because there is a serious discrepancy between the two reports in the
depth range 5--25 m, which contributes significantly to the seismic
gravity gradients.  The Skagit velocities there are more plausible
than the LIGO ones\footnote{The report prepared for LIGO
{\cite{dames_moore}} claims $c_P=1400$ m/s, $c_S=370$ m/s,
corresponding to a Poisson ratio of $\nu=0.46$.  This could be
appropriate for water-saturated materials at this depth, but is not
appropriate for the dry materials that actually lie there.  The Skagit
report {\cite{skagit}} shows two layers in this range of depths: one
with $c_P=520$ m/s, $c_S=270$ m/s, for which $\nu=0.32$; the other
with $c_P =820$ m/s, $c_S=460$ m/s, for which $\nu=0.27$.  For dry
alluvia, these values are much more reasonable than $\nu=0.46.$ We
thank Alan Rohay for bring this point to our attention.}.

Table {\ref{table:skagit}} shows velocity profiles as extracted from
the Skagit report. Notice the overall gradual increase in both wave
speeds.  This is due to compression of the alluvia by the weight of
overlying material, with a consequent increase in the areas of the
contact surfaces between adjacent particles (silt, sand, or gravel)
\cite{ishihara}.  Notice also the sudden increase of $c_P$ and $\nu$
at 40 m depth, due to a transition from dry alluvia to
water-saturation; the water contributes to the bulk modulus but not
the shear modulus, and thence to $c_P$ but not $c_S$.  Notice,
finally, the large jump in both $c_P$ and $c_S$ at the 220 m deep
transition from alluvial deposits to basalt.

\begin{table}
\caption{Velocity profiles at the Hanford LIGO site, as extracted from Table
2.5--3, Fig. 2.5--10, and Sec.\ 2.5.2.5 of the Skagit Report
{\protect\cite{skagit}}.  These velocities are based on (i) cross-hole
measurements (waves excited in one borehole and measured in another)
down to 60 m depth; (ii) downhole measurements (waves excited at
surface and arrivals measured in boreholes) from $z=60$ m to $z=175$
m; (iii) extrapolations of downhole measurements at other nearby
locations, and surface refraction measurements (waves excited at
surface and measured at surface) at the LIGO site, from $z=175$ m down
into the basalt at $z>220$ m.  The downhole measurements at one well
(Rattlesnake Hills No.\ 1) have gone into the basalt to a depth of
$3230$ m.  Depths are in meters, velocities are in m/s.
\label{table:skagit}}
\vskip 15pt
\begin{tabular}{llll}
Depths&$c_P$&$c_S$&$\nu$\\
\tableline
0--12&520&270&0.32\\
12--24&820&460&0.27\\
24--32&1000&520&0.31\\
32--40&1260&530&0.39\\
40--50&1980&560&0.46\\
50--80&2700&760&0.46\\
80--110&2700&910&0.44\\
110--160&1800&610&0.44\\
160--210&2400&910&0.42\\
210--220&2900&1200&0.40\\
220--250&4900&2700&0.28\\
250--3230&5000--5700 competent &&\\
&$\quad\quad\quad\quad\;\;\;$basalt flows &&\\
&4000--5500 interbeds&&\\
\end{tabular}
\end{table}

We have been warned by geophysicist and seismic engineer colleagues
that we should not place great faith in all the details of measured
velocity profiles such as this one; and the discrepancies between the
Skagit and LIGO velocity-profile measurements have reinforced this
caution.  As a result, from computations based on these velocity
profiles (and similar profiles at Livingston), we can only expect to
learn (i) the general nature of the modes to be expected at each LIGO
site, (ii) how those modes' characteristics are influenced by the
velocity profiles, (iii) the range of anisotropy ratios $\cal A$ and
reduced transfer functions $\beta'$ to be expected at each site, and
(iv) how $\cal A$ and $\beta'$ depend on the velocity profiles and the
modes' characteristics.  We {\it cannot} expect the computed,
mode-by-mode details of ${\cal A}(f)$ and $\beta'(f)$ to be accurate
--- except, perhaps, for the shallowly seated RF mode.  Nevertheless,
the insights that we {\it do} gain from such computations should be of
considerable help in future studies of seismic gravity-gradient noise
and future attempts (if any) to mitigate it.

In this spirit, we have simplified our calculations by approximating
the measured Hanford velocity profiles (Table {\ref{table:skagit}})
with their twelve distinct layers by the simpler four-layer model
shown in Table {\ref{table:4LayerHanford}}.  Layers 1 and 2 are dry
alluvia, layer 3 is water-saturated alluvium, and layer 4 is basalt.

\begin{table}
\caption{Four-layer model for the velocity profiles at the Hanford LIGO site.
Notation: $n$ --- layer number, $D_n$ --- layer thickness, $c_{Pn}$
--- P-wave speed in this layer, $c_{Sn}$ --- S-wave speed in this
layer, $\nu_n$ --- Poisson ratio in this layer.  Depths and
thicknesses are in meters, speeds are in m/s.
\label{table:4LayerHanford}}
\vskip15pt
\begin{tabular}{llllll}
$n$&Depths&$D_n$&$c_{Pn}$&$c_{Sn}$&$\nu_n$\\
\tableline
1&0--12&12&520&270&0.32\\
2&12--40&28&900&500&0.28\\
3&40--220&180&2400&700&0.45\\
4&220--4000&$3780$&4900&2700&0.28\\
\end{tabular}
\end{table}

\subsection{Hanford model results}
\label{subsec:HanfordResults}

The horizontally stratified geologies at Hanford and Livingston
support a variety of Love and Rayleigh modes.  (For the general
character of Love and Rayleigh modes see, {\it e.g.}, Refs.\
{\cite{pilant,eringen}} and the brief discussion in the introduction
of this paper.)  We shall focus on Rayleigh modes in this section,
since they are the sole producers of seismic gravity-gradient noise.

In each geological layer, consider a specific Rayleigh mode.  It
consists of a superposition of plane-fronted P- and SV-waves.  Because
each layer is idealized as homogeneous, the mode's SV- and P-waves are
decoupled within the layer. However, they are coupled at layer
interfaces and the earth's surface by the requirement that material
displacement and normal stress be continuous across the interface (or
with the atmosphere in the case of the earth's surface).  The details
of this coupling and its consequences are worked out in Appendix
{\ref{app:Multilayer}}.

In each layer, the mode's P- and SV-components propagate at different
angles to the vertical: $\alpha_{Pn}$ for the P-waves in layer $n$ and
$\alpha_{Sn}$ for the SV-waves.  However, the components must all move
with the same horizontal speed
\begin{equation}
c_H = {\omega\over k} =
{c_{Pn}\over \sin\alpha_{Pn} } = {c_{Sn}\over \sin\alpha_{Sn}}
\label{eq:Snell}
\end{equation}
(Snell's law), and they must all have the same horizontal wave number
$k$ and frequency $\omega=2\pi f$.

Each mode can be characterized by its dispersion relation for
horizontal motion $\omega(k)$, or equivalently $c_H(f)$.  It will be
helpful, in sorting out the properties of the modes, to understand
first what their dispersion relations $c_H(f)$ would be if their
SV-wave components were decoupled from their P-wave components.  We
shall do so in the next subsection, and then examine the effects of
coupling in the following subsection.  Note that we shall ignore the
effects of damping in these two subsections, since the lengthscales
involved are less than (or at most of the same order as) the
dissipation lengthscales of both P- and SV-waves [{\it cf.}\ Eq.\
({\ref{eq:attenuation}})].

\subsubsection{P-SV decoupling approximation}
\label{sec:PSVdecoupling}

Recall that we denote by RP$n$ the $n$th Rayleigh mode of P-type and
by RS$n$ the $n$th Rayleigh mode of SV-type.  In the approximation of
P-SV decoupling, Mode RP$n$ with horizontal speed $c_H$ propagates
from the earth's surface through sequences of strata (generating no
SV-waves) until it reaches a depth $D_P$ where $c_P$ first exceeds
$c_H$.  At that location, it reflects and returns to the surface, and
then is reflected back downward.  The mode's dispersion relation
$c_H(f)$ is determined by the resonance condition that the reflected
waves arrive at the surface in phase with the original downgoing
waves.

\begin{figure}
\epsfxsize=3.3in\epsfbox{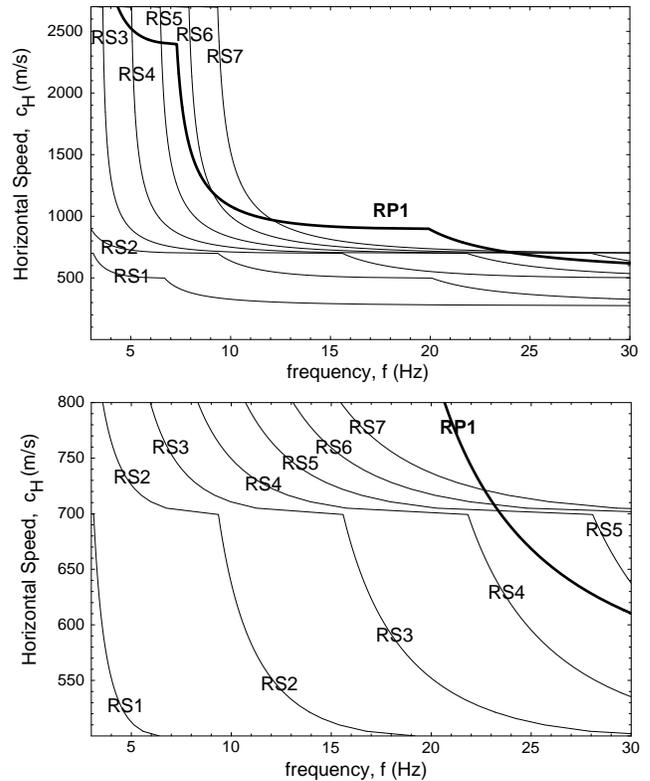}
\caption{
Dispersion relations for the 4-layer Hanford model, as computed using the
P-SV decoupling approximation, Eqs.\ (\protect\ref{eq:DRPGeomOpt}) and
(\protect\ref{eq:DRSGeomOpt}).
}
\label{fig:GeomOptDRHanford}
\end{figure}

This resonance condition is evaluated most easily by following the
(locally) planar waves vertically downward to their reflection point
(the location $z=D_P$ where $c_P$ first reaches $c_H$) and then back
up, thereby returning precisely to the starting point.  On this path,
the vertical component of the wave vector is
\begin{equation}
k_V = {\omega\over c_P} \cos\alpha_P = {\omega\over c_P} \sqrt{1-(c_P/c_H)^2}
 = k \cot \alpha_P\;,
\label{eq:kV}
\end{equation}
where we have used Snell's law (\ref{eq:Snell}) to infer $\cos\alpha_P
= \sqrt{1-(c_P/c_H)^2}$.  The waves' corresponding waves' total
round-trip phase shift is
\begin{equation}
{\Delta\Phi} = 2 \int_0^{D_{P}} {\omega\over c_P}
\sqrt{1-(c_P/c_H)^2}\, dz + {\delta\Phi}_{\rm interfaces}\;.
\label{eq:DeltaPhi}
\end{equation}
Here $\delta\Phi_{\rm interfaces}$ is the total phase shift acquired
at the interfaces between strata and upon reflecting at the earth's
surface.  Setting $\omega = 2\pi f$ and imposing the resonance
condition $\Delta\Phi = 2n\pi$, we obtain the following dispersion
relation for mode RP$n$:
\begin{equation}
f={n-(\delta\Phi_{\rm interfaces}/2\pi) \over 2 \int_0^{D_P} \sqrt{
c_P^{-2} - c_H^{-2} }dz}\;.
\label{eq:DRPGeomOpt}
\end{equation}
Similarly, for mode RS$n$ the dispersion relation is
\begin{equation}
f={n-(\delta\Phi_{\rm interfaces}/2\pi) \over 2 \int_0^{D_S} \sqrt{ c_S^{-2} -
c_H^{-2} }dz}\;,
\label{eq:DRSGeomOpt}
\end{equation}
where $D_S$ is the depth at which $c_S$ first reaches $c_H$.

Figure {\ref{fig:GeomOptDRHanford}} shows these
decoupling-approximation dispersion relations for our 4-layer model of
$c_P(z)$ and $c_S(z)$ (Table {\ref{table:4LayerHanford}}).  For the
RS-waves, the total interface phase shift has been set to
$\delta\Phi_{\rm interfaces} = \pi$, which would be the value for a
single layer with a huge rise of $c_S$ at its base.  For the sole RP
mode shown, RP1, it has been set to $\delta\Phi_{\rm interfaces} =
\pi/2$, which is a fit to the dispersion relation with P-SV coupling
(Fig.\ \ref{fig:DRHanford}, to be discussed below).

Notice that for fixed horizontal speed $c_H$, the lowest RP mode, RP1,
occurs at a much higher frequency $f$ than the lowest RS mode, RS1.
This is because of the disparity in propagation speeds, $c_P =
\mbox{several} \times c_S$.  Notice also the long, flat plateaus in
$c_H(f)$ near $c_H=c_{S2}=500$ m/s and especially $c_{S3}=700$ m/s for
the RS$n$ modes, and near $c_H=c_{P2}=900$ m/s and $c_H=c_{P3}=2400$
m/s for RP1.  Mathematically these are caused by the vanishing square
roots in the denominators of the dispersion relations
(\ref{eq:DRPGeomOpt}) and (\ref{eq:DRSGeomOpt}).  Physically they
arise because the mode's waves ``like'' to propagate horizontally in
their deepest layer.  At high frequencies ({\it e.g.}, $f\agt 10$ Hz
for $c_H \simeq c_{S3} = 700$ m/s), several modes propagate together
nearly horizontally in that deepest layer.

\subsubsection{Effects of P-SV coupling on dispersion relations}

\begin{figure}
\epsfxsize=3.3in\epsfbox{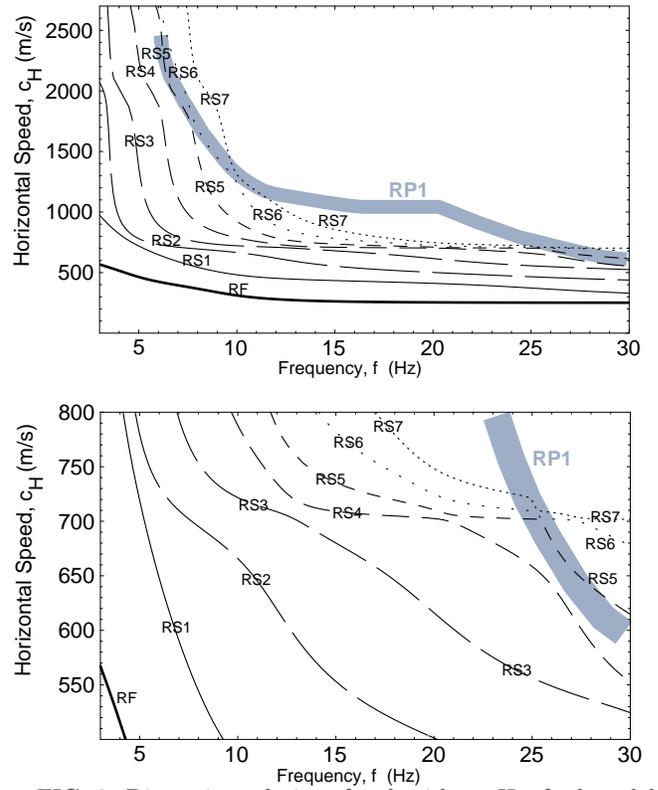}
\caption
{Dispersion relations for the 4-layer Hanford model, including coupling
between P- and SV-waves produced at boundaries between layers and at the
earth's surface.}
\label{fig:DRHanford}
\end{figure}

Figure {\ref{fig:DRHanford}} shows the dispersion relations $c_H(f)$
for the lowest 8 modes of our 4-layer model at Hanford, with P-SV
coupling included.  These dispersion relations were computed using the
multilayer equations of Appendix {\ref{app:Multilayer}}.  We shall now
discuss these various dispersion relations, beginning with that for
the fundamental mode, which is labeled RF in the figure.

Mode RF was studied in Sec.\ {\ref{subsec:HHRayleigh}} for an
idealized homogeneous half space.  It is vertically evanescent in both
its P- and SV-components (except at low frequencies in the top layer);
for this reason, it did not show up in our idealized
decoupling-approximation dispersion relation (Fig.\
{\ref{fig:GeomOptDRHanford}}).  At frequencies $f\agt 10$ Hz, its
vertical $e$-folding lengths ${\cal Z}_P$ and ${\cal Z}_S$ [Eqs.\
(\ref{eq:LHH})] are both short enough that it hardly feels the
interface between layers 1 and 2, and the homogeneous-half-space
description is rather good.  Below 10 Hz, interaction with the
interface and with layer 2 pushes $c_H$ up.

By contrast with the P-SV-decoupled Fig.\
{\ref{fig:GeomOptDRHanford}}, every Rayleigh overtone mode RP$n$ or
RS$n$ in Fig.\ \ref{fig:DRHanford} now contains a mixture of SV- and
P-waves.  This mixture varies with depth in the strata and is
generated by the same kind of interface reflection and refraction as
we met in Sec.\ {\ref{subsec:PUpSVUp}} for SV-up and P-up waves.  In
most regions of the $(c_H,f)$ plane, the mode mixtures are dominated
either by SV- or P-waves --- the ratio of energy in one wave type to
that in the other is $>2$.

In the vicinity of the wide gray band marked RP1, the modes are
predominately of RP type; away from that vicinity they are
predominately RS.  The location of the RP1 band has been inferred from
the computed S- and P-wave amplitudes.  Notice how well it agrees with
the decoupling approximation's RP1 dispersion relation (Fig.\
{\ref{fig:GeomOptDRHanford}}).  Away from the RP1 band, the dispersion
relation for each RS$n$ mode is reasonably close to its
decoupling-approximation form (compare Figs.\ {\ref{fig:DRHanford}}
and {\ref{fig:GeomOptDRHanford}}).  As each mode nears and crosses the
RP1 band, its dispersion relation is distorted to approximately
coincide, for awhile, with the RP1 shape.  Correspondingly, all its
other properties become, for awhile, those of an RP mode.

\subsubsection{Anisotropy ratios and reduced transfer functions}
\label{subsec:AbetapHanford}

\begin{figure}
\epsfxsize=3.3in\epsfbox{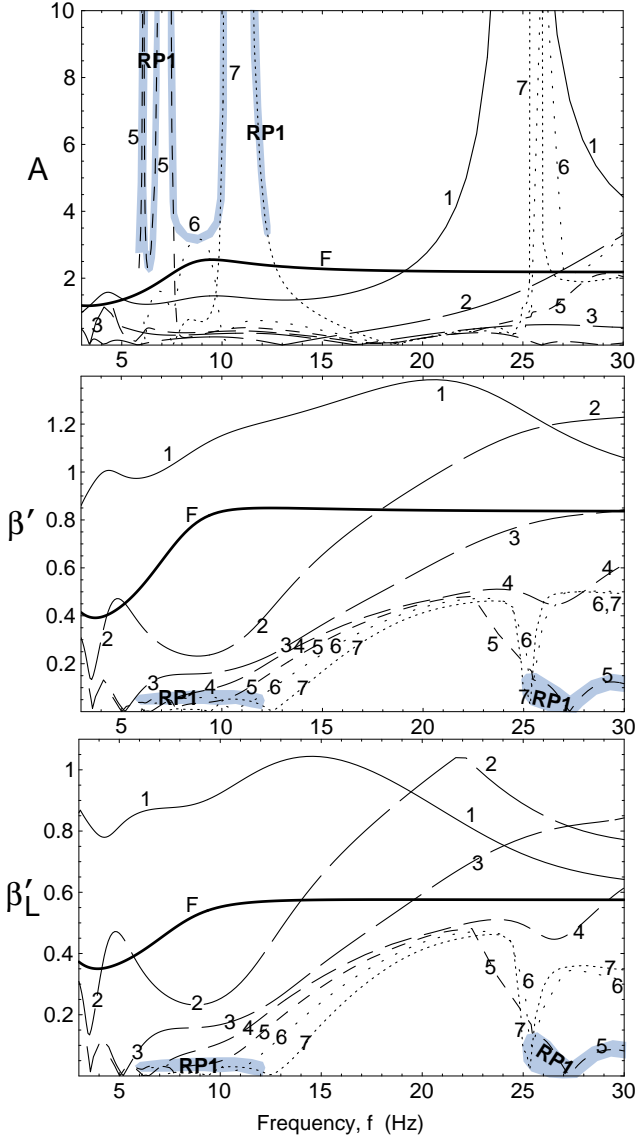}
\caption
{Properties of the lowest 8 modes of the 4-layer Hanford model,
including coupling between P- and SV-waves produced at boundaries
between layers and at the earth's surface.}
\label{fig:PropertiesHanford}
\end{figure}

Figure {\ref{fig:PropertiesHanford}} shows the anisotropy ratio $\cal
A$ and reduced transfer function $\beta'$ for the lowest eight modes
of our 4-layer model of Hanford.  These were computed using the
multilayer equations of Appendix {\ref{app:Multilayer}}, with
dissipation neglected.  On the Figure, the mode names ``RS$n$'' have
been shortened to ``$n$'', and ``RF'' to ``F''.  The bottom set of
graphs is the value $\beta'_L$ that the net reduced transfer function
would have if the mode of interest were mixed with enough Love waves
to reduce the net anisotropy ratio to the value ${\cal A}_{\rm net}
\simeq 1.0$ typical of measured seismic spectra at Hanford during
quiet times {\cite{rohayhanf}}.

{\bf Fundamental Mode RF.}  Above 10 Hz, mode RF has ${\cal A} \simeq
2.2$, $\beta'\simeq 0.84$, and $\beta'_L \simeq 0.58$, in accord with
our homogeneous-half-space model (Sec.\
{\ref{sec:HomogeneousHalfSpace}}).  Below 10 Hz, coupling of the RF
mode to layer 2 produces a growth of the subsurface source to
partially cancel the surface source, and a resulting fall of $\beta'$
to 0.4 and $\beta'_L$ to 0.35.

{\bf RS Overtones.}  In RS regions (away from the RP1 band) the
overtone modes RS$n$ generally have ${\cal A}\alt 1$ so
$\beta'_L\simeq\beta'$ --- little or no admixed Love waves are needed
to bring the anisotropy down to 1.0.  The value of $\beta'$ ranges
from $\sim 0.4$ to $1.4$ in the RS regions; but when the RP1 mode is
nearby in the $c_H$--$f$ plane, its admixture drives $\beta'$ down to
$\alt0.2$.

Mode RS1 shows characteristic ``SV-up'' behavior near 25 Hz (compare
Fig.\ {\ref{fig:PropertiesHanford}} with Figs.\ {\ref{fig:AnisoUp}}
and {\ref{fig:betapUp}}).  Its $\cal A$ has a very large resonance and
its subsurface source (not shown in the figures) has a sharp dip to
nearly zero, resulting from $45^\circ$ upward propagation of its
SV-component in the top layer and no production of P-waves upon
reflection.  At frequencies above our range of interest, this same
SV-Up behavior will occur in successively higher RS$n$ modes.

{\bf RP1 Mode.}  The region of RP1 behavior is shown as thick gray
bands in Fig.\ {\ref{fig:PropertiesHanford}} ({\it cf.}\ the bands in
Fig.\ {\ref{fig:DRHanford}}).  The RP1 reduced transfer function is
small, $\alt 0.15$, due to the same near-cancellation of its surface
and subsurface sources as we met for P-Up waves in Sec.\
\ref{subsec:PUpSVUp} and Fig.\
\ref{fig:betapUp}. As each RS mode crosses the core
of the RP1 region, its $\beta'$ shows a dip and its
anisotropy shows a peak, revealing the temporary transition to RP
behavior. 

{\bf Higher-order RP Modes.}  The higher-order RP modes ($n=2,3,...$)
in our frequency band will lie in the vicinity of RS$n$ overtones with
$n >8$.  We expect these RP$n$ modes to show similarly small reduced
transfer functions to those for RP1, but we have not attempted to
compute them, with one important exception: high-order RP modes that
travel nearly horizontally in Hanford's $\sim 4$ km thick basalt
layer.  We consider these modes in the next subsection.

\subsubsection{RP modes that travel horizontally in the basalt}
\label{subsec:RPBasalt}

As discussed in the Introduction (Sec.\ {\ref{subsec:OurModes}}), the
ground motions at the Hanford corner and end stations sometimes show
time delays in correlated motion, corresponding to wave propagation
speeds of $\sim 5000$--$6000$ m/s {\cite{rohayhanf,rohaydiff}}.  These
motions must be due to wave modes that travel nearly horizontally in
the $\sim 4$ km thick basalt layer at the base of the alluvium, or in
the bedrock beneath the basalt.  We have computed the properties of
such wave modes for the case of horizontal propagation in the basalt
layer --- layer 4 of our 4-layer Hanford model.

Because of the many closely spaced modes in the relevant $(c_H,f)$
region ($c_H$ a little larger than $c_{P4} = 4900$m/s, $3\,{\rm Hz}\le
f\le 30\,{\rm Hz}$), it is not reasonable, or even of interest, to
compute their dispersion relations explicitly.  Instead, we have
assumed an idealized dispersion relation $c_H = 4910$ m/s independent
of frequency.

The basalt layer is so thick that nearly horizontally propagating
waves will be substantially damped in traveling from its lower face to
its upper face and back; and, the S-waves will be much more strongly
damped than the P-waves.  For this reason, we idealize these waves as
purely P-up as they impinge from the basalt layer 4 onto the layer
3--4 interface.  These P-up waves at interface 3--4 are treated as a
source for other wave components in all 4 layers.

\begin{figure}
\epsfxsize=3.3in\epsfbox{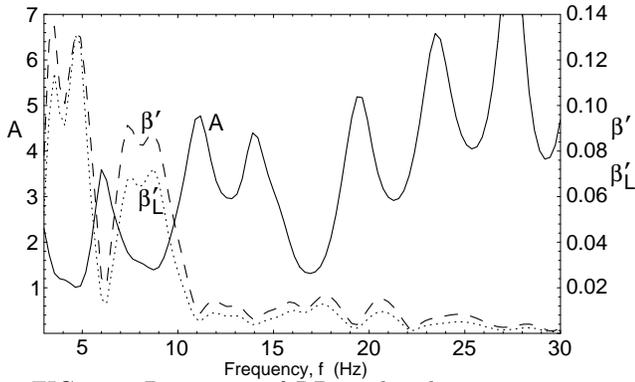}
\caption{
Properties of RP modes that propagate nearly horizontally in layer 4
($\sim 4$ km thick basalt layer) at Hanford ($c_H$ slightly larger
than $c_{P4}=4900$ m/s), including the effects of dissipation in the
alluvium above the basalt.  }
\label{fig:RPHHanford}
\end{figure}

For these waves, dissipation [Eqs.\ (\ref{eq:attenuation}) and
associated discussion] may be more important than for the RF, RS and
RP1 modes, which were treated above as dissipationless.  We therefore
include it in our analysis. We do so in the 4-layer equations of
Appendix {\ref{app:Multilayer}} by giving the sound speeds appropriate
imaginary parts,
\begin{eqnarray}
{\Im(c_{Pn}) \over \Re(c_{Pn})} &=& -{1\over 2Q_P} = - 0.015\;, \nonumber\\
{\Im(c_{Sn}) \over \Re(c_{Sn})} &=& -{1\over 2Q_S} = - 0.03\;,
\label{eq:ImC}
\end{eqnarray}
while keeping their real parts equal to the values shown in Table
{\ref{table:4LayerHanford}}.  We have solved the resulting multilayer
equations numerically, obtaining the anisotropy ratios and reduced
transfer functions shown in Fig.\ {\ref{fig:RPHHanford}}.

The peaks in $\cal A$ at $f\simeq 3$, 11, 19, and 27 Hz [frequency
separation $\Delta f = c_{P3}/\left(2D_3\sqrt{1-c_{P3}^2/c_H^2}\right)
= 8$ Hz] are associated with resonant P-wave excitations of layer 3
and their influence on layer 2 and thence on layer 1; {\it cf.}\ the
decoupling-approximation dispersion relation (\ref{eq:DRPGeomOpt}).
The slightly smaller peaks at $f\simeq 6$, 14, and 23 Hz are due to
resonant S-wave excitations of layer 2.  The oscillations in both
$\cal A$ and (less obviously) $\beta'$ with frequency $\Delta f =
c_{P2}/\left(2D_2\sqrt{1-c_{P2}^2/c_H^2}\right) = 16$ Hz are
associated with resonant P-wave excitations of layer 2.

Despite the complexity of these deeply seated RP modes, with various
types of resonant excitations of various layers, and despite the fact
that the seismic gravity gradients arise from depths ${\cal Z}_{\rm
sgg} = 1/k \sim 25$ to 250 m so great that the top three layers all
contribute, these modes exhibit the same range of values of $\beta'$
as mode RP1: 0 to 0.15.  Here, as for RP1 and for P-up waves in a
homogeneous half space, $\beta'$ is small because of near cancellation
of the gravity gradients from the P-wave surface and subsurface
sources.

\subsubsection{Summary of Hanford model results}

The most important of the above results are those for the reduced
transfer functions of the various modes at Hanford.  They are
summarized in Table {\ref{table:betap}} and their implications are
discussed in the Introduction, Sec.\ {\ref{subsec:OurModes}}.

\section{Livingston}
\label{sec:Livingston}

\subsection{Livingston geophysical structure}

At the LIGO site near Livingston, Louisiana, the geological strata
consist of alluvial deposits laid down by water flowing into the Gulf
of Mexico.  As the ocean level has risen and fallen, alluvial terraces
of varying thickness have been formed.  This alluvium (layers of clay,
silt, sand, and gravel in various orders) is of the Holocene,
Pleistocene, and Pliocene eras going down to a depth of about $700$ m,
and compacted alluvium of the upper Miocene and earlier eras below
that.  These sedimentary deposits extend down to a depth of about $3$
km {\cite{CRC}} before reaching bedrock.

For our analysis the principal issue is the vertical velocity profiles
$c_P(z)$ and $c_S(z)$.  The primary difference between Livingston and
Hanford is the depth of the water table: it is only about $2$ m down
at Livingston, versus about $40$ m at Hanford.  This difference should
cause $c_P$ to soar to about $1600$ m/s at depths of a few meters at
Livingston; it only does so roughly $40$ m down at Hanford.

The only measurements of the Livingston velocity profiles that we have
been able to find are those performed in a site survey for LIGO
{\cite{woodward_clyde}}.  Those measurements only include $c_S$, not
$c_P$, and only go down to a depth of 15 m.  Accordingly, we have had
to estimate the velocity profiles from these sparse data and from the
lore accumulated by the geophysics and seismic engineering
communities.

That lore suggests that $c_S$ should increase as about the $1/4$ power
of depth {\cite{ishihara}}.  (This increase is due to the fact that
the shear restoring force must be carried by the small-area interfaces
between the grains of gravel, sand, silt, or clay; the weight of
overlying material compacts the grains, increasing the areas of their
interfaces.)  We have fit the measured $c_S(z)$ in the top 15 m ({\{7
ft, 700 ft/s\}}, {\{21 ft, 810 ft/s\}}, {\{50 ft, 960 ft/s\}}) (Ref.\
{\cite{woodward_clyde}, Appendix B, plate 7) to a 1/4 power law,
adjusting the fit somewhat to give speeds at greater depths in rough
accord with measurements at a similar sedimentary site in Tennessee
{\cite{tennessee}}.  Our resulting fit is
\begin{mathletters}
\label{eq:cScPLFit}
\begin{equation}
c_S = 185\,{\rm m/s}(1+z/2.9\,{\rm m})^{1/4}\;.
\label{eq:cSLFit}
\end{equation}
A combination of theory and phenomenology [Eqs.\ (6.24), (6.26) of
{\cite{ishihara}} and associated discussion] tells us that in these
water-saturated alluvia, the material's Poisson ratio should be about
\begin{equation}
\nu = {1\over2}\left[1-0.39 \left({c_S\over1000\,{\rm m}}\right)^2\right]\;.
\label{eq:nuL}
\end{equation}
(The Poisson ratio goes down gradually with increasing compaction and
increasing $c_S$ because water is playing a decreasing role compared
to the grains.)  The standard relation
\begin{equation}
c_P = c_S\sqrt{ 2-2\nu\over1-2\nu}\;,
\label{eq:cPL}
\end{equation}
\end{mathletters}
combined with Eqs.\ (\ref{eq:cSLFit}) and (\ref{eq:nuL}), then gives us the
vertical profile for $c_P$.

These profiles are valid only in the water-saturated region.  Although
the water table is at $\sim 2$ m, measurements elsewhere
{\cite{tennessee}} suggest that one may have to go downward an
additional several meters before the effects of the water on $c_P$
will be fully felt.  Accordingly, we expect $c_P \sim 2 c_S$ in the
top $\sim 5$ m at Livingston, followed by a sharp rise to the values
dictated by Eqs.\ (\ref{eq:cSLFit})--(\ref{eq:cPL}), though in our
final conclusions (Sec.\ \ref{subsec:LivingstonRFRS}), we shall allow
for the possibility that the sharp rise occurs at anywhere from 2 to 5
m depth.

\subsection{Livingston 4-layer model}

We have fit a four-layer model to these estimated Livingston velocity
profiles.  Our fit is shown in Table \ref{table:4LayerLivingston}.
This model is the primary foundation for our exploration of seismic
gravity gradients at Livingston. As discussed above, it principally
differs from the 4-layer Hanford model by the rapid increase of $c_P$
at 5 m depth at Livingston, due to the higher water table.  All other
differences have a much more minor influence on the seismic
gravity-gradient noise.
\begin{table}
\caption{Four-layer model for the velocity profiles at
the Livingston LIGO site.
Notation and units are as in Table
{\protect\ref{table:4LayerHanford}}.
\label{table:4LayerLivingston}}
\vskip15pt
\begin{tabular}{llllll}
$n$&Depths&$D_n$&$c_{Pn}$&$c_{Sn}$&$\nu_n$\\
\tableline
1&0--5&5&440&220&0.33\\
2&5--105&100&1660&400&0.47\\
3&105--905&800&1700&700&0.40\\
4&905--3005&2100&1900&1000&0.31\\
\end{tabular}
\end{table}

\subsection{Livingston model results}

\subsubsection{Mode overview}

Because the top, unsaturated layer is so thin, RP modes cannot
resonate in it in our frequency band; and because water makes $c_P$ so
large just below the top layer, the RP modes in our band can only
propagate at a correspondingly high speed, $c_H > 1660$ m/s.  The
lowest 10 RS modes, by contrast, are confined to speeds $c_H\alt1000$
m/s.  As a result --- in contrast to Hanford --- there is no mixing
between these lowest RS modes and the RP modes.  The RS modes have
purely RS character, with no significant RP admixture.

In the next section we shall study the lowest 10 RS modes along with
the fundamental mode.  In the following section, we shall examine the
lowest RP mode.

\subsubsection{RF and RS modes}
\label{subsec:LivingstonRFRS}

We have computed the dispersion relations, anisotropy ratios, and
reduced transfer functions for modes RF and RS1--10 in our 4-layer
Livingston model, using the multilayer equations of Appendix
{\ref{app:Multilayer}}.  The dispersion relations are shown in Fig.\
{\ref{fig:DRLivingston}}.  Because of the separation in the $(c_H,f)$
plane of these modes from the RP mode, we expect the P-SV decoupling
approximation to work quite well here.  Indeed, the RS modes have the
form one would expect from the decoupling approximation [Eqs.\
(\ref{eq:DRSGeomOpt})].  The anisotropy ratios and reduced transfer
functions are shown in Fig.\ {\ref{fig:PropertiesLivingston}}.

\begin{figure}
\epsfxsize=3.3in\epsfbox{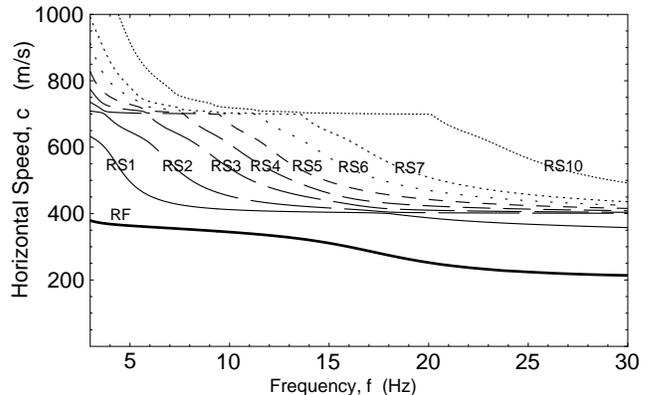}
\caption{
Dispersion relations for 4-layer Livingston model,
including coupling between P- and SV-waves produced at boundaries
between layers, for the fundamental mode and the lowest 10 RS modes.}
\label{fig:DRLivingston}
\end{figure}

\begin{figure}
\epsfxsize=3.3in\epsfbox{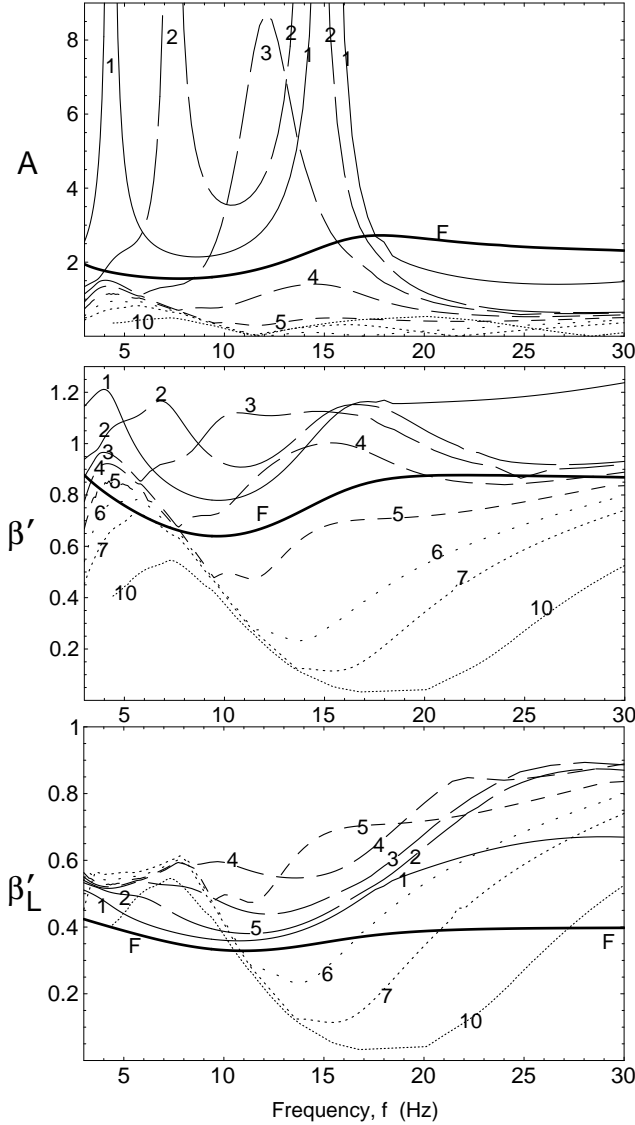}
\caption{
Properties of the lowest 10 RS modes and the RF mode of the 4-layer
Livingston model.  (Modes RS8 and RS9 are not shown; their curves are
sandwiched between 7 and 10.)}
\label{fig:PropertiesLivingston}
\end{figure}

{\bf RF Mode.}  Because the top layer is $2.5$ times thinner in our
Livingston model than at Hanford, the frequency at which the RF mode
becomes like that of a homogeneous half space is $2.5$ times higher:
$\sim25$ Hz compared to $\sim10$ Hz.  Only above $\sim 25$ Hz do the
mode's properties asymptote toward their homogeneous-half-space values
of $c_H = 205$ m/s, ${\cal A} = 2.2$, $\beta'=0.86$ and $\beta'_L =
0.40$.  At lower frequencies, interaction with layer 2 pushes $\beta'$
into the range 0.65--0.9, and $\beta'_L$ into the range 0.35--0.45.

It is possible that the effects of water saturation will cause $c_P$
to shoot up at depths shallower than the $5$ m assumed in our model; a
transition anywhere in the range $2\,{\rm m}\alt z\alt5\,{\rm m}$ must
be considered reasonable.  If the transition in fact occurs at depths
shallower than $5$ m, the peaks of $\beta'$ and $\beta'_L$ will be
pushed to correspondingly higher frequencies.  Thus, we must be
prepared for the RF mode to have $\beta'$ anywhere in the range
$0.65$--$0.9$, and $\beta'_L$ in the range $0.35$--$0.45$ at just
about any frequency in our band of interest.

{\bf RS Modes.}  In our frequency band, the RS modes have negligible
excitation in layers 3 and 4, and their P-waves are evanescent in
layers 2, 3 and 4.  As a result, these modes can be well approximated
by SV-up waves in layer 2, impinging on the layer 1--2 interface.  We
have verified this by computing their anisotropies and reduced
transfer functions in this 2-layer SV-up approximation by the method
outlined at the beginning of Appendix {\ref{app:PupSup}}.  The results
for ${\cal A}$ and $\beta'$, which relied on the 4-layer dispersion
relations of Fig.\ {\ref{fig:DRLivingston}}, agree to within a few per
cent with those of our 4-layer model (Fig.\
{\ref{fig:PropertiesLivingston}}) except at frequencies below $5$ Hz
where the differences become somewhat larger.

Throughout our frequency band these RS modes have vertical
seismic-gravity-gradient $e$-folding lengths ${\cal Z}_{\rm sgg} = 1/k
\agt D_1 = 5$ m.  Thus, the upper parts of layer 2 contribute
significantly to the reduced transfer function $\beta'$, along with
all of layer 1.

For modes RS1--RS5, the gravity gradients are largely due to the
S-waves' vertical surface motions, and correspondingly the reduced
transfer functions have the familiar range $\beta' \simeq 0.6$--1.2
that we encountered for RS modes at Hanford (Sec.\
\ref{subsec:AbetapHanford}) and for SV-Up modes in a homogeneous half
space (Fig.\ \ref{fig:betapUp}).

By contrast, modes RS6--RS10 show a phenomenon not exhibited at
Hanford: a broad dip in $\beta'$ to a value $\ll 1$.  This dip is
caused by a significant excitation of P-waves in Livingston's 5-meter
thick top layer: the vertical surface motions in the dip are largely
due to the P-waves, and mass conservation guarantees that the gravity
gradients they produce will be nearly cancelled by those from the
subsurface, P-wave compressional source.  These surface-layer
excitations are {\em not} associated with any RP mode; as we shall see
in the next subsection, the lowest RP mode at these frequencies has
$c_H$ about twice as high as for these modes.  It seems that the close
proximity of the two very sharp geophysical discontinuities (the
earth's surface and the sharp rise of $c_P$ caused by water) forces
the modes' S-waves to generate a sizable component of P-waves even
moderately far from P-wave resonance.  No such phenomenon was observed
in our Hanford 4-layer model.

\subsubsection{Mode RP1}
\label{sec:LRP1}

Figure {\ref{fig:DRLivingstonP}} shows the dispersion relation for the
lowest RP mode, RP1, at Livingston, along with the RF and lowest 10 RS
modes.  As noted earlier, RP1 does not overlap the other modes [by
contrast with Hanford (Figs.\ {\ref{fig:GeomOptDRHanford}} and
{\ref{fig:DRHanford}})].

\begin{figure}
\epsfxsize=3.3in\epsfbox{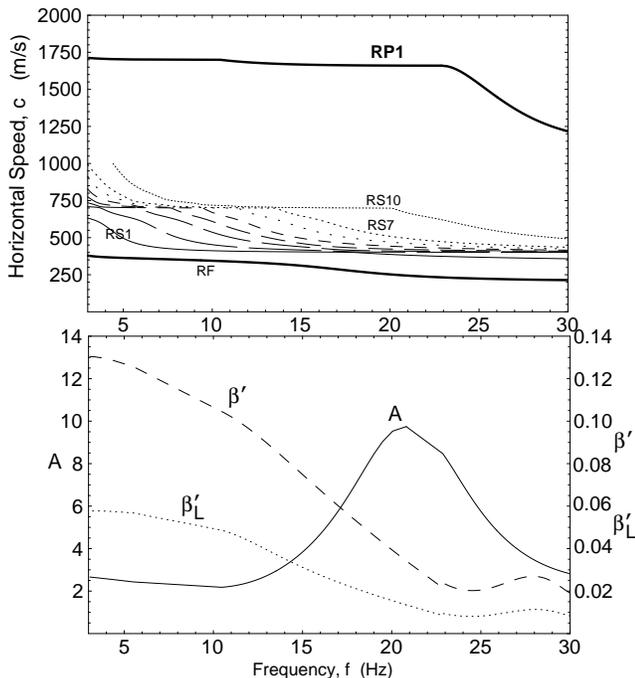}
\caption{
Upper panel: dispersion relation for mode RP1 in the 4-layer
Livingston model.  Also shown for comparison are the dispersion
relations for the fundamental Rayleigh mode RF and the lowest few RS
modes ({\it cf.}\ Fig.\ {\protect\ref{fig:DRLivingston}}).  Lower
panel: properties of mode RP1.  }
\label{fig:DRLivingstonP}
\end{figure}

At frequencies $f<22.8$ Hz, the RP1 mode has horizontal speed
$c_H>c_{P2} = 1660$ m/s and thus its P-waves can propagate in layers 1
and 2 (and also in layer 3 below $11.3$ Hz).  In this region we have
evaluated $c_H(f)$ using the P-SV decoupling approximation [Eq.\
(\ref{eq:DRPGeomOpt})].

At frequencies $f>22.8$ Hz, the horizontal speed is $c_H < c_{P2}$, so
the mode's P-waves are evanescent in layers 2, 3, and 4.  In this
regime we have adopted an approximation that is much more accurate
than the decoupling one.  We have idealized the material as
two-layered: a 5 m thick upper layer with the properties of layer 1 of
Table {\ref{table:4LayerLivingston}}, and below that a homogeneous
half space with the properties of layer 2.  For this
layer-plus-half-space model we have used an analytic dispersion
relation due to Lee {\cite{lee}} (Appendix {\ref{app:TwoLayer}}).
Because the mode's SV-waves can leak out of layer 1 into layer 2 (and
then propagate away to ``infinity'' --- or, more realistically,
dissipate), Lee's dispersion relation predicts a complex frequency $f$
if $c_H$ is chosen real, and a complex $c_H$ if $f$ is chosen real.
The predicted losses are small (quality factors $Q$ decreasing from
$\simeq 50$ at $f\simeq24$ Hz to $\simeq 15$ at $30$ Hz).  The real
part of the dispersion relation is shown in the upper panel of Fig.\
{\ref{fig:DRLivingstonP}}.

The lower panel of Fig.\ {\ref{fig:DRLivingstonP}} shows the
anisotropies and reduced transfer functions for this RP1 mode.  At
$f<22.8$ Hz, where the P-waves are propagating nearly horizontally in
layer 2, these properties were computed using the P-up approximation
in the above two-layer (layer-plus-half-space) model; {\it cf.}\ the
introduction to Appendix {\ref{app:PupSup}}. More specifically, the
dispersion relation (with both $c_H$ and $f$ real) was taken from the
P-SV decoupling approximation, the P-waves with this $c_H$ and $f$
were regarded as impinging from layer 2 onto the top of layer 1 at a
glancing angle, and the reflected P- and SV-waves were regarded as
propagating off to ``infinity'' (or, more realistically, dissipating
before any return to the interface).  This is the approximation that
was so successful for the RS modes when combined with the correct
4-layer dispersion relation, but we don't have a good handle on its
accuracy here, with the less reliable P-SV-decoupling dispersion
relation.  We are much more confident of our approximation for
$f>22.8$ Hz.  There we used the exact two-layer equations (Appendix
{\ref{app:Multilayer}}), together with Lee's exact, complex dispersion
relation $c_H(f)$.

These computations produced an anisotropy that peaks at $f=22.8$ Hz
where $c_H=c_{P2}$, with a peak value of ${\cal A} \sim 8$ (Fig.\
{\ref{fig:DRLivingstonP}}).  This is smaller than the peak
anisotropies for mode RP1 at Hanford (Fig.\
{\ref{fig:PropertiesHanford}}), but comparable to those for the
higher-order RP modes that propagate nearly horizontally in the
Hanford basalt (Fig.\ {\ref{fig:RPHHanford}}).

The reduced transfer function $\beta'$ lies in the same range, 0 to
0.15, as for all the Hanford RP modes that we studied (Figs.\
{\ref{fig:PropertiesHanford}} and {\ref{fig:RPHHanford}}).  This adds
to our conviction that this low range of $\beta'$ is a general
characteristic of RP modes.

\subsubsection{Summary of Livingston model results}

The most important of the above results are those for the reduced
transfer functions $\beta'$ of the various modes at Livingston.  They
are summarized in Table {\ref{table:betap}} and their implications are
discussed in the Introduction, Sec.\ {\ref{subsec:OurModes}}.

\section{Concluding Remarks}
\label{sec:Conclusions}

\subsection{Summary}

In this paper, we have used the theory of seismic surface waves to
calculate the seismic gravity-gradient noise spectra that are to be
expected at the Hanford, Washington and Livingston, Louisiana LIGO
sites.  Our final noise strengths, as shown in Fig.\ {\ref{fig:hSGG}},
are close to Saulson's previous rough estimate.  At noisy times and
near 10 Hz, the seismic gravity-gradient noise is likely to be more
serious than vibrational and thermal seismic noise in advanced
interferometers.  Unless means are found to combat gravity-gradient
noise (see below for possible methods), the hard-won gains in
sensitivity due to R\&D on vibration isolation and thermal noise may
be compromised by seismic gravity gradients, at least at noisy times.

\subsection{Effects of topography and of LIGO construction}

In our analysis we have idealized the earth's surface near the LIGO
test masses as perfectly planar and as undisturbed by LIGO
construction.  Irregularities in topography will significantly disturb
the waves' propagation and their vertical structure only if the
surface height varies by amounts as large as $\sim2\,{\rm m}/(f/10{\rm
Hz}) = (\sim1/2$ the shortest vertical $e$-folding length ${\cal Z}_P$
for RF waves), on horizontal lengthscales as short as $\sim8\,{\rm
m}/(f/10{\rm Hz}) = (\sim2$ times the horizontal reduced wavelength
$1/k$ of those RF waves), within distances of the test masses $\sim
25\,{\rm m}/(f/10{\rm Hz}) = ($the horizontal wavelength of those RF
waves), for frequencies $\sim 3-30$ Hz.  (Of all the modes we have
studied, the RF modes hug the surface most tightly and thus will be
most influenced by the topography.)

Variations on these scales were rare at the two LIGO sites before
construction.  However, the grading that made the arms flat produced
topographic variations in the vicinity of some of the test masses that
are marginally large enough to disturb the propagation. Examples are
the long pits dug alongside the arms at Livingston to get material for
building up the arms' heights, and excavation at Hanford to lower the
arms below the level of the surrounding land near the southwest arm's
midstation and the northwest arm's endstation.

We speculate that these topographic modifications will alter the
seismic gravity gradient noise by a few tens of percent, but probably
not by as much as a factor 2.  Future studies should examine this
issue.

The 1 m deep concrete foundations of the buildings that house the test
masses will likely also influence the noise by a few tens of percent,
particularly at $\sim 20-30$ Hz where the RF waves' vertical
penetration is short.  The foundation extends approximately $10$
meters by $25$ meters at the interferometer's end stations (and also,
in the case of Hanford, at the mid station).  The foundation is
approximately ``X'' shaped for the corner stations, with each arm of
the ``X'' extending roughly $100$ meters by $20$ meters
{\cite{blueprints}}.  The sound speeds in the concrete will be a
factor of several higher than the surrounding ground, so the
foundations will form very sharp ``geophysical'' interfaces in the
ground, causing diffraction of impinging waves and altering their
vertical structure.  Because the foundations are so shallow, we doubt
that their net effect on the seismic gravity gradient noise can be as
large as a factor 2, but future studies should examine it.

\subsection{Measurements that could firm up our understanding
of seismic gravity gradients}

Our analysis is plagued by a large number of uncertainties regarding
the true make-up of the ambient seismic background at the LIGO sites.
We made extensive use of measurements of ground motion which
functioned as constraints on what modes could be present.  These
measurements were helpful, but certain other measurements would be
considerably more helpful.  We suggest that, to the extent that
resources permit, these measurements be included in future seismic
surveys for gravitational-wave interferometer sites, including future
surveys at the LIGO sites.

First, we recommend careful measurements of the sound speeds and
dynamical Poisson ratios of the ground as a function of depth,
especially in the top few tens of meters and if possible down to the
bedrock.  At Hanford, we had reasonably complete data {\cite{skagit}},
thanks to earlier plans to build a nuclear power plant in the
vicinity.  As discussed in this paper, we encountered serious
discrepancies between those old data and data from the LIGO
geotechnical survey.  At Livingston, we had no P-wave speed or Poisson
ratio profiles, and the S-wave speed profiles available only went down
to a depth of 15 meters.  As a result, we had to use a mixture of
theory, profiles from other sites, and phenomenological fitting to
obtain a plausible velocity profile.  Velocity profiles are of crucial
importance in determining how the various modes behave in the ground.

Second, we recommend measurements that more nearly directly determine
the modes that characterize the seismic motion.  In this paper, as
discussed above, we were able to put together very rough estimates of
the modes that actually characterize the seismic background by using
surface motion data as constraints, particularly anisotropy ratios
measured at the sites, and by appealing to more detailed measurements
at other sites.  However, other techniques could provide much more
useful and restrictive constraints, thereby more sharply
differentiating among the various modes.  In particular:

\begin{itemize}

\item
Surface seismic arrays {\cite{douze,malagnini}} allow one to measure
the phase relationships of ground motion at appropriately separated
points, from which one can infer the excited modes' wave numbers
$k(f)$ and horizontal propagation speeds $c_H (f)$.

\item
Borehole measurements {\cite{douze}} allow one to measure the phase
correlation of motion at the surface and at some depth $z$
underground, and the variation of amplitudes with depth, thereby
introducing additional constraints on the background.

\item
Specialty seismic instruments called ``dilatometers''
{\cite{hiroo,dilatometers}} measure directly the fractional density
perturbation $\delta\rho/\rho$ that is the subsurface source of
seismic gravity gradients.  Measurements down boreholes with such
devices could place further constraints on the mode mixtures present,
and could show how $\delta\rho/\rho$ varies with depth, at fixed
frequency.  When correlated with vertical surface seismic
measurements, they could give information about the cancellation of
gravity gradients from the surface and subsurface sources.

\end{itemize}

\subsection{Mitigation of seismic gravity gradient noise}

Seismic gravity gradients are unlikely to be a major concern to LIGO
detectors in the near future, since these detectors are only sensitive
to frequencies $f \agt 35$ Hz.  Eventually, however, LIGO
experimenters may succeed in achieving extremely good vibration
isolation and thermal noise control at frequencies $f\alt 10$ Hz.  At
this time, the detectors may well be plagued by seismic
gravity-gradient noise, at least at noisy times; and there may be a
strong need to try to mitigate it.

We see two possibilities for modest amounts of mitigation: (i)
monitoring the noise and removing it from the LIGO data, and (ii)
building moats to impede the propagation of RF-mode seismic waves into
the vicinities of the test masses.

{\bf Monitoring and correction:} By using dedicated 3-dimensional
arrays of vertical surface seismometers and borehole-mounted
dilatometers in the vicinities of all test masses, one might be able
to determine both the surface and subsurface components of
$\delta\rho/\rho$ with sufficient spatial and temporal resolution for
computing the seismic gravity gradient noise and then removing it from
the data.  This paper's insights into the modes at the two LIGO sites
and the gravity gradients they produce may provide a foundation for
future explorations of monitoring-and-correction strategies.

{\bf Moats:} By constructing a narrow, evacuated moat around each test
mass, one might succeed in shielding out a significant portion of the
RF waves that we suspect are the dominant source of quiet-time seismic
gravity gradients.  Since the RF mode contains substantial S-waves and
they are the dominant contributors to the gravity-gradient noise, such
moats may have to be at least as deep as the S-waves' vertical
e-folding length, ${\cal Z}_S \simeq \hbox{9--15} {\rm m} (10 {\rm
Hz}/f)$ [Eq.\ (\ref{eq:LHH}) modified for an increase in the RF speed
$c_H$ due to stratification as shown in Figs. \ref{fig:DRHanford} and
\ref{fig:DRLivingston}].   Since ${\cal Z_S} \simeq 2.5 {\cal Z_P}$, 
moats of this depth would strongly shield out the RF mode's P-waves.

The radius of the moats should be $\agt \lambda \sim \hbox{20--35}
{\rm m} (10{\rm Hz}/f)$.  It is not clear to us whether such moats at
Livingston would be effective if filled with water, or whether they
would have to be kept pumped out.  The water would shield out the RF
mode's S-waves but transmit its P-waves.  If, after transmission, the
waves remain mostly of P-type, then a significant reduction of
$\beta'$ could result; but it is not at all obvious how much
regeneration of S-waves would occur in the moat-surrounded cavity.
Detailed modeling would be required to sort out such issues.

Although moats may be well-suited to reduce gravity gradients
generated by the RF mode, they are probably not so well-suited to
reduce gravity gradients generated by Rayleigh overtones.  The
overtones can be visualized as seismic waves that propagate by
bouncing between layer interfaces and the earth's surface; they could
propagate right under the moat and into the region under the test
mass.  Conceivably, they could even resonantly ``ring'' the earth
under the mass, {\it worsening} the seismic gravity-gradient noise.

If seismic gravity gradients become a problem in the future, ideas
such as moats and monitoring-and-correcting will have to explored.

\section*{Acknowledgments}

We thank Peter Saulson for triggering this research, and Kenneth
Libbrecht, Rai Weiss and Stan Whitcomb for helpful comments.  We thank
geophysicists Hiroo Kanamori and Susan Hough and seismic engineer
Ronald Scott for very helpful conversations and advice, and Alan Rohay
for advice and for providing his measurements of the seismic ground
motion at Hanford and Livingston.  We thank Albert Lazzarini for
facilitating access to Rohay's data sets and to blueprints of the LIGO
site facilities, and Fred Asiri for helping us to track down
information about the geological structures at the sites (including
the Skagit report).  Finally, we thank Giancarlo Cella for providing
us, shortly before this paper was submitted, a copy of the
VIRGO-Project manuscript on seismic gravity gradient noise
{\cite{cella_cuoco}} and for a helpful discussion.  This research was
supported by NSF Grant PHY--9424337.  S.\ A.\ H.\ gratefully
acknowledges the support of the National Science Foundation Graduate
Fellowship Program.  K.\ S.\ T.\ thanks the Max-Planck-Institut f\"ur
Gravitationsphysik for hospitality during the final weeks of writing
this manuscript.

\appendix

\section{General Expression for Reduced Transfer Function}
\label{app:betap}

In this Appendix we derive Eqs.\
(\ref{eq:betaJsplit})--(\ref{eq:betapJ}) for the reduced transfer
function and anisotropy ratio of an arbitrary Rayleigh mode.  In the
text the mode is labeled $J$; in this Appendix we shall omit the
subscript $J$.

The mode has frequency $f$, angular frequency $\omega = 2\pi f$,
horizontal wave number $k$, horizontal phase speed $c_H = \omega/k$,
and horizontal propagation direction $\hat k$.  At the earth's surface
its displacement vector is
\begin{equation}
\vec\xi(z=0)=
(\xi_H \hat k - \xi_V \vec e_z) e^{i(\vec k \cdot \vec x - \omega t)}
\label{eq:eigenvector}
\end{equation}
[Eq.\ (\ref{eq:eigenfunction})]; and on and beneath the surface it produces a
fractional density perturbation
\begin{equation}
{\delta\rho\over\rho} = [\xi_V\delta(z) + {\cal R}(z)]e^{i(\vec k \cdot \vec x
- \omega t)}\;
\label{eq:drhoOverrho1}
\end{equation}
[Eq.\ (\ref{eq:drhoOverrho})];
here $\vec k = k \hat k$ is the horizontal wave vector and $\delta(z)$ is the
Dirac delta function.

Since the ambient seismic motions are horizontally isotropic, this
mode is excited equally strongly for all horizontal directions $\hat
k$, and also for all wave numbers in some (arbitrarily chosen) small
band $\Delta k$ around $k$---{\it i.e.}, in the annulus ${\cal
C}_{\Delta k}$ of width $\Delta k$ in wave-vector space.
Correspondingly (with an arbitrary choice for the strength of the
excitation), the net displacement along some horizontal direction
$\hat n$, in the frequency band $\Delta f = c_H \Delta k / 2\pi$, is
\begin{equation}
X(t) = \Re \left[ \sum_{\vec k} \xi_H (\hat k \cdot \hat n) e^{i(\vec k
\cdot \vec x - \omega t)} \right]\;,
\label{eq:Xt}
\end{equation}
and the power of this random process $X(t)$ in the frequency band $\Delta f$ is 
\begin{equation}
\tilde X^2(f) \Delta f = \sum_{\vec k} |\xi_H|^2 (\hat k \cdot \hat n)^2 =
|\xi_H|^2 {N_{\rm \Delta k}\over 2}   \;, 
\label{eq:Xf}
\end{equation}
where $N_{\rm \Delta k} = \sum_{\vec k} 1$ is the
(normalization-dependent) total number of allowed $\vec k$ values in
the annulus ${\cal C}_{\Delta k}$, and the $1/2$ comes from averaging
$(\hat k\cdot \hat n)^2$ over the horizontal direction $\hat k$.
(Note: the overall normalization $N_{\Delta k}$ of our procedure for
going from the random process expressed as a sum over directions to
the processes's power will have no influence on our final answers for
$\cal A$ and $\beta'$, since they are square roots of ratios of powers
from which $N_{\Delta k}$ drops out.)  Similarly, the net displacement
and power along the vertical $\vec e_z$ direction are
\begin{equation}
Z(t) = \Re \left[- \sum_{\vec k} \xi_V  
e^{i(\vec k \cdot \vec x - \omega t)}\right] \;,
\label{eq:Zt}
\end{equation}
and
\begin{equation}
\tilde Z^2(f) \Delta f = \sum_{\vec k} |\xi_V|^2 =
|\xi_V|^2 N_{\rm \Delta k} \;.
\label{eq:Xf1}
\end{equation}
The mode's anisotropy ratio, ${\cal A} = \tilde Z / \tilde X$ is
therefore
\begin{equation}
{\cal A} = \sqrt2 |\xi_V| / |\xi_H |\;,
\label{eq:alpha1}
\end{equation}
{\it cf.}\ Eq.\ (\ref{eq:calAJ1});
and the direction-averaged power $\tilde W^2\Delta f = (2\tilde X^2 \Delta f
+ \tilde Z^2 \Delta f)/3$ is
\begin{equation}
\tilde W^2 \Delta f = {|\xi_H|^2 + |\xi_V|^2 \over 3} N_{\Delta k}\;.
\label{eq:W2}
\end{equation}

By analogy with Eq.\ (\ref{eq:Xt}), the isotropically excited mode
produces a fractional perturbation in density on and beneath the
earth's surface given by
\begin{equation}
{\delta\rho\over\rho} = 
\Re \left[ \sum_k \left[\xi_V \delta(z) + {\cal R}\right]
e^{i(\vec k \cdot \vec x - \omega t)} \right]\;;
\label{eq:deltarho}
\end{equation}
cf.\ Eq.\ (\ref{eq:drhoOverrho1}).  As an aid in computing the
gravitational acceleration produced on one of the interferometer's
test masses by these density perturbations, we place the origin of
coordinates (temporarily) on the earth's surface, immediately beneath
the test mass. Then the location of the test mass is $-{\cal H}\vec
e_z$, where ${\cal H}$ is its height above the surface.  We denote by
$\hat m$ the unit vector along the laser beam that is monitoring the
test mass's position.  Then the gravitational acceleration along the
$\hat m$ direction is
\begin{equation}
a_{\hat m}(t) = -\int d^3x' {(\vec x' \cdot \hat m) G \delta\rho(\vec x',t) 
\over |\vec x' + {\cal H}\vec e_z|^3}\;. 
\label{eq:am}
\end{equation}
Invoking Eq.\ (\ref{eq:deltarho}) and introducing Cartesian
coordinates $(x',y',z')$ inside the sum with $\vec k$ along the
$x'$-direction, we bring Eq.\ (\ref{eq:am}) into the form
\begin{eqnarray}
&&a_{\hat m} = -\sum_{\vec k} e^{-i\omega t} G\rho \nonumber \\ 
&&\times \int\!\int\!\int\!{(x' m_x + y' m_y)
e^{ikx'}[\xi_V\delta(z') +{\cal R}(z')]
\over[x'^2 + y'^2 + (z'+{\cal H})^2]^{3/2} }
dz' dx'dy'.
\nonumber\\
\label{eq:am1}
\end{eqnarray}
Integrating out the horizontal directions $x'$ and $y'$ from $-\infty$
to $+\infty$ at fixed $z'$, and integrating out the $\delta$ function,
we obtain our final expression for the gravitational acceleration on
the test mass
\begin{eqnarray}
a_{\hat m} = -\sum_{\vec k} && 2\pi i G \rho (\hat m \cdot \hat k) 
e^{-i\omega t} e^{-k{\cal H}}\nonumber \\
&&
\times \left(\xi_V + \int_0^\infty {\cal R}(z') e^{-kz'}dz'\right)
\;.
\label{eq:am2}
\end{eqnarray}

We next solve the pendular equation of motion for the displacement
$\delta \vec x_j \cdot \hat m_j$ of the test mass in response to this
gravitational acceleration (where the label $j=$ 1, 2, 3, or 4
indicates which of the interferometer's four test masses we are
discussing); the result is
\begin{eqnarray}
\delta \vec x_{j} \cdot \hat m_j = -\sum_{\vec k}&& 
{2\pi i G\rho (\hat k \cdot \hat m_j) e^{i(\vec k
\cdot {\vec x}_j - \omega t)} e^{-k{\cal H}}
\over \omega_0^2 - \omega^2 -i\omega/\tau} \nonumber\\
&&\times \left(\xi_V + \int_0^\infty {\cal R}(z') e^{-kz'}dz'\right)
\;.
\label{eq:delta_x}
\end{eqnarray}
Here $\omega_0$ and $\tau$ are the angular eigenfrequency and damping
time of the test mass's pendular motion.  After completing the
calculation we have moved the origin of coordinates to the
interferometer's beam splitter, thereby producing the term $i \vec k
\cdot \vec x_j$ in the exponential, where $\vec x_j$ is the test
mass's location; {\it cf.}\ Fig.\ {\ref{fig:interferometer}}.

\begin{figure}
\epsfxsize=3in\epsfbox{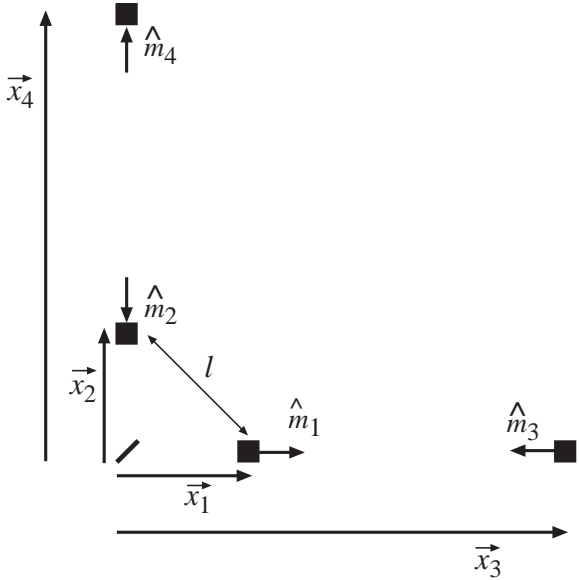}
\caption{The geometry of the interferometer.}
\label{fig:interferometer}
\end{figure}

The interferometer's displacement signal $x(t) = L h(t)$ is its
difference in arm lengths,
\begin{equation}
x(t) = \sum_{j=1}^4 \delta \vec x_j \cdot \hat m_j\;.
\label{eq:xt}
\end{equation}
We have chosen $\hat m_j$ to point away from the test mass's mirror on
the first arm and toward the mirror on the second arm as shown in
Fig.\ {\ref{fig:interferometer}}.  The seismic gravity-gradient noise
is obtained by inserting expression (\ref{eq:delta_x}) into
(\ref{eq:xt}) for each of the four test masses.

The contributions to this noise coming from the two end masses, $j=$ 3
and 4, are not correlated with those coming from any other test mass
in our 3--30 Hz frequency band, since 3 and 4 are each so far from the
corner and each other ($4\,{\rm km} \gg \lambda = 2\pi/k$).  However,
there is a significant correlation between the two corner test masses,
1 and 2.  Taking account of this correlation, the interferometer's
displacement signal $x(t)$ [Eqs.\ (\ref{eq:delta_x}) and
(\ref{eq:xt})] exhibits the following noise power in the frequency
band $\Delta f$:
\begin{eqnarray}
\tilde x^2(f) \Delta f &=& {(2\pi G\rho)^2\over (\omega^2 -
\omega_0^2)^2 + \omega^2/\tau^2} e^{-2k{\cal H}}\nonumber\\
&&\times\sum_{\Delta k} 
\left|\xi_V + \int_0^\infty {\cal R}(z') e^{-kz'}dz'\right|^2 J_k\;,
\label{eq:xf}
\end{eqnarray}
where 
\begin{eqnarray}
J_k &=& \sum_{\hat k} \left[ |\hat k \cdot\hat m_1 e^{i\vec k\cdot\vec x_1} +
\hat k \cdot \hat m_2 e^{i\vec k\cdot \vec x_2}|^2 \right. \nonumber\\
&&\left. \;\;\;\;\;\;\;+ (\hat k \cdot \hat m_3)^2 +
(\hat k \cdot \hat m_4)^2 \right]\;.
\label{eq:Jk}
\end{eqnarray}
Here we have broken up the sum over $\vec k$ into one over all
directions $\hat k$ and one over its length $k$ in the range $\Delta
k$.  Each of the last two terms in $J_k$ (the uncorrelated
contributions of masses 3 and 4) average to 1/2, and the first term
can be rewritten in terms of $\vec x_1 - \vec x_2$:
\begin{equation}
J_k = \sum_{\hat k} \left[ |\hat k \cdot \hat m_1 e^{i\vec k\cdot(\vec
x_1 - \vec x_2)} + \hat k \cdot \hat m_2 |^2 + 1 \right] 
\label{eq:Jk1}
\end{equation}
By virtue of the geometry of the interferometer's corner test masses
(Fig.\ {\ref{fig:interferometer}}), ${\vec x}_1 - {\vec x}_2 = l({\hat
m}_1 + {\hat m}_2)/\sqrt2$, where $l$ is the separation between those
masses.  Inserting this into Eq.\ (\ref{eq:Jk1}), setting ${\hat
k}{\cdot}{\hat m}_1 = \cos\phi$ and ${\hat k}{\cdot}{\hat m}_2 =
{\sin\phi}$, and averaging the quantity inside the sum over ${\hat k}$
({\it i.e.}, over $\phi$), we obtain
\begin{equation}
J_k = 2\sum_{\hat k} \gamma^2(kl) \;,
\label{eq:Jk2}
\end{equation}
where 
\begin{equation}
\gamma(y) =\sqrt{1+ {1\over2\pi} \int_0^{2\pi} \!\!\!\!\cos\phi \sin\phi
\cos\left(y {\cos\phi + \sin\phi \over \sqrt2} \right)d\phi}\;
\label{eq:gamma1}
\end{equation}
[Eq.\ (\ref{eq:gamma})].  This function is graphed in Fig.\
{\ref{fig:gamma}}.  Inserting Eq.\ (\ref{eq:Jk2}) into Eq.\
(\ref{eq:xf}) and noting that $\sum_{\Delta k}\sum_{\hat k} =
\sum_{\vec k} = N_{\Delta k}$ is the number of allowed wave vectors in
the annulus ${\cal C}_{\Delta k}$, we obtain our final expression for
the interferometer's displacement noise power:
\begin{eqnarray}
\tilde x^2(f) \Delta f &=& {(4\pi G\rho)^2\over (\omega^2 -
\omega_0^2)^2 + \omega^2/\tau^2}\,
\gamma^2\left(\omega l \over c_H \right) e^{-2k{\cal H}}   \nonumber\\
&&
\quad\times \left|\xi_V+\int_0^\infty {\cal R}(z') e^{-kz'} dz' \right|^2 
{N_{\Delta k}\over 2}\;.
\label{eq:xf2}
\end{eqnarray}

The transfer function $T(f)$ for the seismic gravity-gradient noise is
obtained by dividing the direction-averaged ground displacement noise
power (\ref{eq:W2}) into the interferometer displacement noise power
(\ref{eq:xf2}) and taking the square root.  The result is expression
(\ref{eq:Tbeta}) with the reduced transfer function $\beta$ given by
$\beta = \gamma \Gamma \beta'$ [Eq.\ (\ref{eq:betaJsplit})], where
$\Gamma = e^{-k{\cal H}}$ [Eq.\ (\ref{eq:GammaJ}) in which
$\omega/c_{H J} = k$] and
\begin{equation}
\beta'(f) = \sqrt{3/2\over |\xi_{H }|^2 + |\xi_{V }|^2}
\left| \xi_V + \int_0^\infty {\cal R}(z')
e^{-kz'} dz'\right|\;;
\label{eq:betap1}
\end{equation}
[Eq.\ (\ref{eq:betapJ})].

\section{Fundamental Rayleigh mode in homogeneous half space}
\label{app:HomogeneousHalfSpace}

\vskip 10pt\noindent
In this Appendix we briefly review the theory of Rayleigh waves
propagating in a homogeneous half space ({\it i.e.}, a homogeneous,
planar model of the earth), and then we derive the anisotropy ratio
$\cal A$ and reduced transfer function $\beta'$ for such waves.

A homogeneous half space can support only the fundamental Rayleigh
mode, since the overtones all require inhomogeneities to confine them
in the vicinity of the earth's surface.  The theory of this mode is
developed in a variety of standard texts
{\cite{landau_lifshitz,love,pilant,eringen}}.  According to that
theory, the waves propagate with a horizontal speed $c_H$ which is
slightly slower than the S-wave speed $c_S$ (which in turn is slower
than $c_P$).  The ratio $c_H/c_S$ is a function of the material's
Poisson ratio $\nu$, varying from $c_H/c_S = 0.904$ for $\nu = 0.16$
(fused quartz) to $c_H/c_S = 0.955$ for $\nu = 0.5$ (fluids and other
easily sheared materials).  More generically, it is given by $c_H/c_S
= \sqrt{\zeta}$, where $\zeta$ is the real root, in the range
$0<\zeta<1$, of the equation
\begin{equation}
\zeta^3 - 8\zeta^2 + 8\left({2-\nu \over 1-\nu}\right) \zeta -
{8\over(1-\nu)} = 0\;.
\label{eq:cR}
\end{equation}
The Rayleigh waves' horizontal wave number is $k = \omega/c_H$, and
their wavelength is $\lambda = 2\pi/k$.  The P-wave of the fundamental
Rayleigh mode decays with depth $z$ as $e^{-qkz}$, where the
dimensionless ratio $q$ of vertical $e$-folding rate to horizontal
wave number is
\begin{equation}
q = \sqrt{1-(c_H/c_P)^2} \;.
\label{eq:q}
\end{equation} 
Similarly, the SV-wave part decays with depth as $e^{-skz}$, where the
dimensionless ratio $s$ of vertical $e$-folding rate to horizontal
wave number is
\begin{equation}
s = \sqrt{1-(c_H/c_S)^2} = \sqrt{1-\zeta} \;.
\label{eq:s}
\end{equation}

More specifically, the mode's displacement eigenvector $\vec \xi$ can
be decomposed into a P-wave which is the gradient of a scalar
potential plus an SV-wave which is the curl of a vector potential.  We
shall denote by $\psi$ the complex amplitude of the scalar potential.
The normal components of elastodynamic stress produced by this wave
must vanish\footnote{More accurately, they must be continuous with the
stress produced by the earth's atmosphere, which we approximate as
vacuum.}  at the earth's surface.  Upon imposing these boundary
conditions, a standard calculation {\cite{landau_lifshitz,love}} gives
the following expression for the displacement vector:
\begin{eqnarray}
\vec\xi =  &&  ik\psi\left( e^{-qkz} - {2qs\over
1+s^2} e^{-skz}\right) e^{i(\vec k \cdot \vec x-\omega t)}
\hat k \nonumber\\
&& -qk\psi\left( e^{-qkz} - {2\over
1+s^2} e^{-skz}\right) e^{i(\vec k \cdot \vec x-\omega t)} \vec e_z
\;. \nonumber\\
\label{eq:xi}
\end{eqnarray}
Here, $\vec e_z$ is the unit vector pointing in the $z$-direction,
which we take to be down, $t$ is time, $\vec x$ denotes horizontal
location, and $\vec k = k \hat k$ is the mode's horizontal wave
vector.  By comparing this displacement vector with Eq.\
(\ref{eq:eigenfunction}), we read off the following expressions for
the horizontal and vertical displacement amplitudes at the earth's
surface, $z=0$:
\begin{equation}
\xi_H = ik\psi\left( {1+s^2-2qs \over 1+s^2}\right)\;, \quad
\xi_V = - qk\psi \left( {1-s^2\over1+s^2}\right)\;. 
\label{eq:RFxiH}
\end{equation}
The wave displacement (\ref{eq:xi}) produces a fractional perturbation
$\delta\rho/\rho$ of the earth's density beneath the surface given by
\begin{equation}
{\delta\rho(z>0)\over\rho} 
= -\vec\nabla\cdot\vec\xi = {\cal R} e^{i(\vec k \cdot
\vec x - \omega t)}\;,
\label{eq:RFdrho}
\end{equation}
where
\begin{equation}
{\cal R}(z) = (1-q^2) k^2\psi e^{-qkz}\;.
\label{eq:RFRz}
\end{equation}

Inserting Eqs.\ (\ref{eq:RFxiH}) into Eq.\ (\ref{eq:calAJ1}), we
obtain the anisotropy ratio for the RF mode of a homogeneous half
space,
\begin{equation}
{\cal A} = \sqrt{2}{q(1-s^2)\over 1+s^2-2qs} \;,
\label{eq:calAR}
\end{equation}
and inserting (\ref{eq:RFxiH}) and (\ref{eq:RFRz}) into
(\ref{eq:betapJ}) and integrating, we obtain the mode's reduced
transfer function
\begin{equation}
\beta' = \sqrt{3(1+s^2-2q)^2\over 2(1+s^2)[(1+s^2)(1+q^2)-4qs]}\;.
\label{eq:betaR}
\end{equation}

\newpage
\widetext
\onecolumn

\section{Multilayer model}
\label{app:Multilayer}

\vskip 10pt\noindent
In this Appendix we derive the equations governing Rayleigh overtones
and the reduced transfer function in a multilayer model of geophysical
strata.

\subsection{Model and notation}

Our model consists of $N$ homogeneous layers labeled by the index
$n=1,2,3,\ldots,N$.  Layer 1 is at the surface, layer $N$ is a
homogeneous half space at the bottom, and the interfaces between
layers are horizontal.  The Rayleigh modes propagate as decoupled
planar SV- and P-waves in each layer; they are coupled at the
interfaces by continuous-displacement and continuous-normal-stress
boundary conditions.

We have already introduced much of our notation in the body of the
paper; to make this Appendix self-contained, we reiterate some of it
here, along with some new notation:
\medskip\par
$\omega = 2\pi f$: Angular frequency of waves.
\smallskip\par
$\vec k = k \hat k$: Horizontal wave vector, with $k$ its magnitude
and $\hat k$ the unit vector in its direction.
\smallskip\par
$c_H = \omega/k$: Horizontal phase velocity of waves.
\smallskip\par
$\vec e_z$: Downward pointing unit vector.
\smallskip\par
$D_n$: Thickness of layer $n$.
\smallskip\par
$z_n$: Depth below the top of layer $n$.
\smallskip\par
$\vec\xi_n$: Displacement vector for waves in layer $n$.
\smallskip\par
$K_n$: Bulk modulus in layer $n$.
\smallskip\par
$\mu_n$: Shear modulus in layer $n$.
\smallskip\par
$\rho_n$: Density in layer $n$.
\smallskip\par
$c_{Pn}$: Speed of propagation of P-waves in layer $n$.
\smallskip\par
$c_{Sn}$: Speed of propagation of S-waves in layer $n$.
\smallskip\par
$\alpha_{Pn}$: Angle to vertical of P-wave propagation direction
(between 0 and $\pi/2$ if real, by convention).  If P-waves are
evanescent in the layer, $\alpha_{Pn}$ will be complex.
\smallskip\par
$\alpha_{Sn}$: Angle to vertical of SV-wave propagation vector
(between 0 and $\pi/2$ if real, by convention).  If SV-waves are
evanescent in the layer, $\alpha_{Sn}$ will be complex.
\smallskip\par
${\cal P}_{n }$: Complex amplitude of upgoing P-waves at the top of
layer $n$.
\smallskip\par
${\cal P'}_{n }$: Complex amplitude of downgoing P-waves 
at the top of layer $n$.
\smallskip\par
${\cal S}_{n }$: Complex amplitude of upgoing SV-waves
at the top of layer $n$.
\smallskip\par
${\cal S'}_{n }$: Complex amplitude of downgoing SV-waves
at the top of layer $n$.
\medskip\par
\noindent
In accord with this notation, the displacement vector in layer $n$ has
the following form:
\begin{eqnarray}
\vec\xi_{n} = e^{i(\vec k \cdot \vec x - \omega t)}
&&\left[\left({\cal P}'_{n } e^{ikz_n\cot\alpha_{Pn}}
+ {\cal P}_{n } e^{-ikz_n\cot\alpha_{Pn}}\right )\sin\alpha_{Pn} \;\hat k
+ \left({\cal P}'_{n } e^{ikz_n\cot\alpha_{Pn}}
- {\cal P}_{n } e^{-ikz_n\cot\alpha_{Pn}}\right )\cos\alpha_{Pn} \; 
\vec e_z\right.
\nonumber \\
&&\left. + \left({\cal S}'_{n } e^{ikz_n\cot\alpha_{Sn}}
- {\cal S}_{n } e^{-ikz_n\cot\alpha_{Sn}}\right )\cos\alpha_{Sn} \; \hat k
- \left({\cal S}'_{n } e^{ikz_n\cot\alpha_{Sn}}
+ {\cal S}_{n } e^{-ikz_n\cot\alpha_{Sn}}\right )\sin\alpha_{Sn} \;
\vec e_z \right]. \nonumber\\
\label{eq:xiPSV}
\end{eqnarray}
Since the waves are generated at the Earth's surface, the upward
propagating waves are absent in the lowermost layer:
\begin{equation}
{\cal P}_{N } = 0\;, \quad {\cal S}_{N } = 0\;.
\label{eq:upvanish}
\end{equation}
Consequently, the waves have $4N-2$ complex amplitudes.

\subsection{Equations for the dispersion relation, the propagation
angles, and the amplitudes}

Once one has specified the Rayleigh mode of interest, its horizontal
propagation direction $\hat k$, and one of its amplitudes, say ${\cal
P}_1$, then all its other properties are uniquely determined as a
function of frequency.  To evaluate its properties one first computes
its horizontal dispersion relation $\omega(k)$ [or equivalently
$c_H(f)$] by a procedure to be outlined below.  Then one computes all
the waves' propagation angles by imposing Snell's law ({\it i.e.}, by
demanding that all components of the wave propagate with the same
horizontal speed $c_H$):
\begin{equation}
{c_{Pn}\over\sin\alpha_{Pn}} = {c_{Sn}\over\sin\alpha_{Sn}} = c_H\;.
\label{eq:snell}
\end{equation}

At the Earth's surface, the (primed) amplitudes of the reflected waves
are related to the (unprimed) amplitudes of the incident waves by the
following two standard equations
{\cite{landau_lifshitz,love,pilant,eringen}}:
\begin{eqnarray}
&&2\sin\alpha_{S1} \cos\alpha_{P1} ({\cal P}'_{1} - {\cal P}_{1}) 
+ \cos 2\alpha_{S1} ( {\cal S}'_{1} + {\cal S}_{1}) = 0
\nonumber \\
&&\sin\alpha_{P1} \cos2\alpha_{S1} ({\cal P}'_{1} + {\cal P}_{1})
- \sin\alpha_{S1} \sin2\alpha_{S1} ( {\cal S}'_{1} - {\cal S}_{1})
\nonumber \\ 
&& \quad \quad = 0\;. 
\label{PSVsurface:hanf}
\end{eqnarray}
These equations can be derived by setting the vertical-vertical and
vertical-horizontal components of the stress to zero at the Earth's
surface, and by expressing the ratio of bulk to shear modulus in terms
of the propagation angles:
\begin{equation}
{K_n\over\mu_n} = {{c_{Pn}}^2 \over {c_{Sn}}^2} - {4\over3} = 
{\sin^2\alpha_{Pn} \over \sin^2\alpha_{Sn}} - {4\over3}\;.
\label{eq:Kovermu}
\end{equation}

The junction conditions at the interface between layer $n$ and layer
$n+1$ take the following form \cite{pilant,eringen}:

\begin{mathletters}
\label{PSVjunction:hanf}
\begin{eqnarray}
&&\left({\cal P}'_{n} e^{ i k D_n\cot\alpha_{Pn} } +
{\cal P}_{n} e^{-i k D_n\cot\alpha_{Pn} }\right)\sin\alpha_{Pn} +
\left({\cal S}'_{n} e^{ i k D_n\cot\alpha_{Sn} } -
{\cal S}_{n} e^{-i k D_n\cot\alpha_{Sn} }\right)\cos\alpha_{Sn}
\nonumber\\
&&\quad=\left({\cal P}'_{n+1} + {\cal P}_{n+1} \right)\sin\alpha_{Pn+1} +
\left({\cal S}'_{n+1} - {\cal S}_{n+1} \right)\cos\alpha_{Sn+1}\;,
\label{PSV:intA}
\end{eqnarray}
\begin{eqnarray}
&&\left({\cal P}'_{n} e^{ i k D_n\cot\alpha_{Pn} } -
{\cal P}_{n} e^{-i k D_n\cot\alpha_{Pn} }\right)\cos\alpha_{Pn} -
\left({\cal S}'_{n} e^{ i k D_n\cot\alpha_{Sn} } +
{\cal S}_{n} e^{-i k D_n\cot\alpha_{Sn} }\right)\sin\alpha_{Sn}
\nonumber\\
&&\quad=\left({\cal P}'_{n+1} - {\cal P}_{n+1} \right)\cos\alpha_{Pn+1} 
- \left({\cal S}'_{n+1} + {\cal S}_{n+1} \right)\sin\alpha_{Sn+1}\;,
\label{PSV:intB}
\end{eqnarray}
\begin{eqnarray}
&&{\mu_n}\left[(1 - \cot^2\alpha_{Sn})
\left({\cal P}'_{n} e^{ i k D_n\cot\alpha_{Pn} }+
{\cal P}_{n} e^{-i k D_n\cot\alpha_{Pn} }\right)\sin\alpha_{Pn}+
2\left({\cal S}'_{n} e^{ i k D_n\cot\alpha_{Sn} }-
{\cal S}_{n} e^{-i k D_n\cot\alpha_{Sn} }\right)\cos\alpha_{Sn}\right]
\nonumber\\
&&\quad= {\mu_{n+1}}\left[(1 - \cot^2\alpha_{Sn+1})
\left({\cal P}'_{n+1} + {\cal P}_{n+1} \right)\sin\alpha_{Pn+1}
+ 2\left({\cal S}'_{n+1} - {\cal S}_{n+1} \right)\cos\alpha_{Sn+1}
\right]\;,
\label{PSV:intC}
\end{eqnarray}
\begin{eqnarray}
&&{\mu_n}
\left[2 \left({\cal P}'_{n} e^{ i k D_n\cot\alpha_{Pn} }-
{\cal P}_{n} e^{-i k D_n\cot\alpha_{Pn} }\right)\cos\alpha_{Pn}-
(1-\cot^2\alpha_{Sn})\left({\cal S}'_{n} e^{ i k D_n\cot\alpha_{Sn} }+
{\cal S}_{n} e^{-i k D_n\cot\alpha_{Sn} }\right)\sin\alpha_{Sn}\right]
\nonumber\\
&&\quad={\mu_{n+1}}
\left[2 \left({\cal P}'_{n+1} - {\cal P}_{n+1} \right)\cos\alpha_{Pn+1}
-(1-\cot^2\alpha_{Sn+1})\left({\cal S}'_{n+1} + {\cal S}_{n+1} 
\right)\sin\alpha_{Sn+1}\right]\;.
\label{PSV:intD}
\end{eqnarray}
\end{mathletters}
\noindent
Equation (\ref{PSV:intA}) is continuity of the horizontal
displacement, (\ref{PSV:intB}) is continuity of the vertical
displacement, (\ref{PSV:intC}) is continuity of the vertical-vertical
component of the stress, and (\ref{PSV:intD}) is continuity of the
vertical-horizontal component of the stress.

Equations (\ref{PSVsurface:hanf}) and
(\ref{PSV:intA})--(\ref{PSV:intD}) are $4N-2$ homogeneous linear
equations for $4N-3$ independent ratios of amplitudes, and for the
horizontal dispersion relation $\omega(k)$ [or equivalently $c_H(f)$].
It is convenient to evaluate the dispersion relation by setting to
zero the determinant of the coefficients of the amplitudes in Eqs.\
(\ref{PSVsurface:hanf}) and (\ref{PSV:intA})--(\ref{PSV:intD}).  The
remaining $4N-3$ amplitudes can then be computed in terms of ${\cal
P}_1$ using any $4N-3$ of these equations.  This was the procedure
used to derive the 4-layer results quoted in the text.  Once the
dispersion relation and the amplitudes have been evaluated as
functions of frequency, the anisotropy ratio and reduced transfer
function can be computed using the equations derived in the following
subsection.

\subsection{Anisotropy ratio, and reduced transfer function}

By comparing Eq.\ (\ref{eq:eigenfunction}) with the displacement
eigenfunction (\ref{eq:xiPSV}) for layer $n=1$, we read off the
horizontal and vertical displacement amplitudes at the earth's
surface:

\begin{eqnarray}
\xi_H &=& ({\cal P}'_{1} + {\cal P}_{1})\sin\alpha_{P1}
+ ({\cal S}'_{1} - {\cal S}_{1})\cos\alpha_{S1} \;, \label{eq:xiHPSV} \\
\xi_V &=&
- ({\cal P}'_{1} - {\cal P}_{1})\cos\alpha_{P1}
+ ({\cal S}'_{1} + {\cal S}_{1})\sin\alpha_{S1}\;. 
\label{eq:xiVPSV}
\end{eqnarray}

The wave displacement (\ref{eq:xiPSV}) produces a fractional density
perturbation $\delta\rho_n/\rho_n = -{\vec\nabla}\cdot{\vec\xi}_n =
{\cal R}_n(z_n) e^{i(\vec k \cdot \vec x - \omega t)}$ in layer $n$,
with amplitude given by
\begin{equation}
{\cal R}_n (z_n) = 
{-ik \over\sin\alpha_{Pn}}
\left( {\cal P}'_{n}e^{ikz_n\cot\alpha_{Pn}}
+{\cal P}_{n}e^{-ikz_n\cot\alpha_{Pn}}\right)\;,
\label{eq:drhoPSV:hanf}
\end{equation}

By inserting Eqs.\ (\ref{eq:xiHPSV}) and (\ref{eq:xiVPSV}) into Eq.\ 
(\ref{eq:calAJ1}), we obtain the anisotropy ratio
\begin{equation}
{\cal A} = \sqrt{2} \left| {({\cal P}'_{1} - {\cal P}_{1})\cos\alpha_{P1}
- ({\cal S}'_{1} + {\cal S}_{1})\sin\alpha_{S1}
\over
({\cal P}'_{1} + {\cal P}_{1})\sin\alpha_{P1}
+ ({\cal S}'_{1} - {\cal S}_{1})\cos\alpha_{S1} } 
\right|\;.
\label{eq:calA_multilayer}
\end{equation}
By inserting Eqs.\ (\ref{eq:xiHPSV}), (\ref{eq:xiVPSV}), 
(\ref{eq:drhoPSV:hanf}), and the relation 
\begin{equation}
z = z_n + \sum_{n'=1}^{n-1} D_{n'}
\end{equation}
into Eq.\ (\ref{eq:betapJ}), integrating, and summing over all four
layers, we obtain the reduced transfer function
\begin{mathletters}
\label{eq:betap4layer}
\begin{equation}
\beta'(f) = {{\cal N}(f)\over{\cal D}(f)}\;,
\label{eq:betap}
\end{equation}
where
\begin{eqnarray}
{\cal N}(f) &=&\sqrt{3\over2}
\Biggl|({\cal P}_1 - {\cal P}'_1) \cos\alpha_P + ({\cal S}_1 + {\cal
S}'_1)\sin\alpha_S \nonumber\\
&&+\sum_{n=1}^N {\rho_n\over\rho_1} \left[
-{\cal P}_{n} e^{i\alpha_{Pn}}
e^{-[k\sum_{n'=1}^{n-1}D_{n'}]} 
\left(1-e^{-[kD_n(1+i\cot\alpha_{P_n})]}\right)\right. \nonumber\\
&&+{\cal P}'_{n} e^{-i\alpha_{P_n}}
e^{-[k\sum_{n'=1}^{n-1}D_{n'}]}
\left.\left(1-e^{-[kD_n(1-i\cot\alpha_{P_n})]}\right)\right]\Biggr|\;,
\label{calN}
\end{eqnarray}
\begin{eqnarray}
{\cal D}^2&&(f) =\Biggl|({\cal P}'_{1} + {\cal P}_{1})
\sin\alpha_{P1} + ({\cal S}'_{1} - {\cal S}_{1})
\cos\alpha_{S1} \Biggr|^2
+ \Biggl| ({\cal P}'_{1} - {\cal P}_{1})\cos\alpha_{P1}
- ({\cal S}'_{1} + {\cal S}_{1})\sin\alpha_{S1} \Biggr|^2\;.
\label{calD}
\end{eqnarray}
\end{mathletters}
In Eq.\ (\ref{calN}) for ${\cal N}(f)$, we have inserted the factor
$\rho_n/\rho_1$ to allow for the possibility (ignored in the text)
that the different layers have different densities.

\newpage
\narrowtext
\twocolumn

\section{Lee's dispersion relation for 2-layer model}
\label{app:TwoLayer}

\vskip 10pt\noindent
When there are only two layers, a top layer with thickness $D$ and a
bottom layer with infinite thickness, the dispersion relation
$\omega(k)$ [or equivalently $c_H(f)$] of the multilayer model
(Appendix {\ref{app:Multilayer}}) can be brought into an explicit form
that permits rapid numerical solutions.  This form was derived by Lee
{\cite{lee}} by manipulating the $6\times6$ determinant of the
coefficients of the amplitudes in Eqs.\ (\ref{PSVsurface:hanf}) and
(\ref{PSV:intA})--(\ref{PSV:intD}).  The standard textbook by Eringen
and {\c S}uhubi \cite{eringen} presents and discusses Lee's dispersion
relation [pages 547--550; note that on the first line of their Eq.\
(7.7.44) $\bar\nu_2$ should be $\bar\nu_1$].  The dispersion relation
consists of the following prescription:

The unknown to be solved for is
\begin{equation}
\zeta = (c_H/c_{S2})^2\;.
\label{eq:zeta}
\end{equation}
At low propagation speeds $c_H$ (high frequencies) the SV-waves in
layer 1 will typically propagate rather than decay, with vertical wave
number divided by horizontal wave number given by
\begin{mathletters}
\label{qsdefs:Hanford}
\begin{eqnarray}
\sigma_1 &=& \sqrt{\zeta(c_{S2}/c_{S1})^2-1} =
\sqrt{(c_H/c_{S1})^2 -1} \nonumber\\ 
&=& \cot\alpha_{S1} \;,
\label{sigma1}
\end{eqnarray}
while the other waves will typically be evanescent with ratios of
$e$-folding rate to horizontal wave number given by
\begin{eqnarray}
q_1 &=& \sqrt{1-\zeta(c_{S2}/c_{P1})^2} =
\sqrt{1 - \left({c_H/c_{P1}}\right)^2},\label{q1}\\
q_2 &=& \sqrt{1-\zeta(c_{S2}/c_{P2})^2} =
\sqrt{1 - \left({c_H/c_{P2}}\right)^2},\label{q2}\\
s_2 &=& \sqrt{1-\zeta} = 
\sqrt{1 - \left({c_H/c_{S2}}\right)^2},\label{s2}
\end{eqnarray}
\end{mathletters}
Regardless of the magnitude of $c_H$ and thence regardless of whether
these quantities are real or imaginary, we regard them all as
functions of $c_H$ given by the above expressions.

We define two quantities
\begin{equation}
Q= \mu_2/\mu_1\;, \quad R = \rho_1/\rho_2\;
\label{eq:QR}
\end{equation}
that appear in what follows.  In terms of $\zeta$, $Q$, and $R$, we
define
\begin{eqnarray}
X &=&Q\zeta - 2(Q-1)\;, \\
Y &=&QR\zeta + 2(Q-1)\;,  \\
Z &=&Q(1-R)\zeta - 2(Q-1)\;, \\
W &=&2(Q-1)\;.
\end{eqnarray}
In this dispersion relation and only here $X,Y,Z,W$ represent these
functions instead of representing earth displacements.  In terms of
the above quantities we define
\begin{eqnarray}
\xi_1 &=& (1-{\sigma_1}^2) \left[ X \cosh(kq_1 D) + {q_2\over q_1}
Y \sinh(kq_1D)\right]
\nonumber\\&+&2\sigma_1\left[q_2 W \sin(k\sigma_1 D) - {1\over\sigma_1}
Z \cos(k\sigma_1 D)\right],
\label{eq:leeDRb} \\
\xi_2 &=& (1-{\sigma_1}^2) \left[ s_2 W \cosh(kq_1 D) + {1\over q_1}Z
\sinh(kq_1 D)\right]
\nonumber\\&+&2\sigma_1\left[X \sin(k\sigma_1 D) - {s_2\over\sigma_1}
Y \cos(k\sigma_1 D)\right],
\label{eq:leeDRc} \\
\eta_1 &=& (1-{\sigma_1}^2) \left[ q_2 W \cos(k\sigma_1 D) +
{1\over \sigma_1} Z \sin(k\sigma_1 D)\right]
\nonumber\\&+&2q_1\left[-X \sinh(kq_1 D) - {q_2\over q_1}
Y \cosh(kq_1 D)\right],
\label{eq:leeDRd} \\
\eta_2 &=& (1-{\sigma_1}^2) \left[ X \cos(k\sigma_1 D) + {s_2\over
\sigma_1} Y \sin(k\sigma_1 D)\right]
\nonumber\\&+&2q_1\left[-s_2 W \sinh(kq_1 D) - {1\over q_1}
Z \cosh(kq_1 D)\right].
\label{eq:leeDRe}
\end{eqnarray}
In terms of these four quantities, Lee's dispersion relation takes the
form
\begin{equation}
F(\zeta,kD) \equiv \xi_1 \eta_2 - \xi_2 \eta_1 = 0\;.
\label{eq:leeDR}
\end{equation}

In the language of Lee's dispersion relation, finding multiple
Rayleigh modes is a matter of finding multiple curves $\zeta(kD)$ that
satisfy (\ref{eq:leeDR}).  Each such $\zeta(kD)$ can be translated
into a corresponding $c_H(f)$, since $c_H = \sqrt{\zeta}c_{S2}$ and
$f=c_H k/2\pi$.  Overtone modes undergo a transition in layer 2 from
propagating and lossy (so that seismic wave energy is lost from layer
1 into layer 2), to evanescent and confined (so the waves are
restricted to the vicinity of the top layer)] at speed $c_H(f) =
c_{S2}$, which is equivalent to $\zeta =1$.  Thus, to produce
dispersion relations for overtone modes, one can look for solutions to
(\ref{eq:leeDR}) in the vicinity of $\zeta=1$, and then, depending on
whether one wants confined modes or lossy modes, trace them from
$\zeta = 1$ to higher frequencies and lower horizontal speeds, or to
lower frequencies and higher horizontal speeds.

In Sec.\ \ref{sec:LRP1} we use Lee's dispersion relation to study the
RP1 mode at Livingston in the lossy regime.

\section{P-up and SV-up modes}
\label{app:PupSup}

\vskip 10pt\noindent
In the text we encounter situations in which one can approximate an
overtone mode as P- or SV-waves that propagate upward through a
homogeneous half space until they encounter the Earth's surface or one
or more layers near the surface, and then (exciting the layers)
reflect back downward with accompanying production of the other type
of wave.  Such ``P-up'' and ``SV-up'' modes can be described by the
multilayer equations of Appendix {\ref{app:Multilayer}}, with the up
(unprimed) amplitudes in the bottom layer (homogeneous half space),
$b$, set to $\{{\cal P}_{b} \neq 0, {\cal S}_{b} = 0\}$ for P-up
modes, and $\{{\cal P}_{b} = 0, {\cal S}_{b} \neq 0\}$ for SV-up
modes.

We can derive simple formulas for the anisotropy ratio $\cal A$ and
reduced transfer function $\beta'$ of such modes for the case of no
surface layers (a pure homogeneous half space):

\subsection{P-up modes in a homogeneous half space}

The displacement function is given by Eq.\ (\ref{eq:xiPSV}) with the
subscript $n$'s deleted since there is only one layer.  The primed
(down) amplitudes are given in terms of the unprimed (up) amplitude
${\cal P}$ by the surface junction conditions (\ref{PSVsurface:hanf});
in particular
\begin{mathletters}
\label{eq:PSVprime}
\begin{eqnarray}
{\cal P}'&=&
{4\cos\alpha_P \sin^3\alpha_S \cos\alpha_S - \sin\alpha_P \cos^2 2\alpha_S
\over
4\cos\alpha_P \sin^3\alpha_S \cos\alpha_S + \sin\alpha_P \cos^2 2\alpha_S}
{\cal P}\;,
\\
{\cal S}' &=&
{4\sin\alpha_P \cos\alpha_P \sin\alpha_S \cos 2\alpha_S
\over
4\cos\alpha_P \sin^3\alpha_S \cos\alpha_S + \sin\alpha_P \cos^2 2\alpha_S}
{\cal P}\;.
\end{eqnarray}
\end{mathletters}
Inserting these into Eq.\ (\ref{eq:calA_multilayer}) we obtain the
following anisotropy ratio:
\begin{equation}
{\cal A} = \sqrt2 \cot 2\alpha_S\;, 
\label{eq:calAP}
\end{equation}
where, by Snell's law [Eq.\ (\ref{eq:Snell})],
\begin{equation}
\alpha_S = \arcsin (c_S/c_H)\;.
\label{eq:alphaS}
\end{equation}
Inserting expressions (\ref{eq:PSVprime}) into the one-layer version
of equations (\ref{eq:betap4layer}), we obtain the following reduced
transfer function:
\begin{equation}
{\beta'} = \sqrt6 \sin^2 \alpha_S \;.
\label{eq:betapP}
\end{equation}
The anisotropies and reduced transfer functions of Eqs.\
(\ref{eq:calAP}) and (\ref{eq:betapP}) are shown graphically in Figs.\
{\ref{fig:AnisoUp}} and {\ref{fig:betapUp}} for $c_P/c_S = 2$
(approximately appropriate to the surface materials at Livingston and
Hanford).

\subsection{SV-up modes in a homogeneous half space}

For SV-up modes, as for P-up modes, the displacement function is given
by Eq.\ (\ref{eq:xiPSV}) with the subscript $n$'s deleted.  The primed
(down) amplitudes are given in terms of the unprimed (up) amplitude
${\cal S}$ by the surface junction conditions (\ref{PSVsurface:hanf});
in particular
\begin{mathletters}
\label{eq:SSVprime}
\begin{eqnarray}
{\cal P}' &=& -{\sin\alpha_S \sin4\alpha_S\over 4\cos\alpha_P
\sin^3\alpha_S \cos\alpha_S + \sin\alpha_P \cos^2 2\alpha_S} {\cal
S}, \\
{\cal S}' &=& {4\cos\alpha_P \sin^3\alpha_S
\cos\alpha_S - \sin\alpha_P \cos^2 2\alpha_S\over
4\cos\alpha_P \sin^3\alpha_S \cos\alpha_S + \sin\alpha_P \cos^2 2\alpha_S}
{\cal S}\;.
\end{eqnarray}
\end{mathletters}
Inserting these into Eq.\ (\ref{eq:calA_multilayer}), we obtain the
following anisotropy ratio
\begin{equation}
{\cal A} = 2\sqrt2 \left|{\cot \alpha_P \over \cot^2\alpha_S-1}\right|\;,
\label{eq:calASV}
\end{equation}
where, by Snell's law,
\begin{equation}
\alpha_S = \arcsin(c_S/c_H)\;, \quad \alpha_P = \arcsin(c_P/c_H)\;.
\label{eq:alphaSP}
\end{equation}
Inserting expressions (\ref{eq:SSVprime}) into the one-layer version
of equations (\ref{eq:betap4layer}), we obtain the following reduced
transfer function:
\begin{equation}
\beta' = {\sqrt{6}\sin^2\alpha_S
\;|1-2i\cot \alpha_P \sin^2\alpha_S \sec 2\alpha_S| 
\over\sqrt{1+(2\;|\cot \alpha_P|\; \sin^2\alpha_S \sec 2\alpha_S)^2 }}\;.
\label{eq:betaSV}
\end{equation}

The anisotropies and reduced transfer functions of Eqs.\
(\ref{eq:calASV}) and (\ref{eq:betaSV}) are shown in Figs.\
{\ref{fig:AnisoUp}} and {\ref{fig:betapUp}}, for $c_P/c_S = 2$.

\end{document}